\documentclass[iop,apj,twocolappendix,numberedappendix]{emulateapj}

\usepackage{rotating}

\usepackage{float}
\floatstyle{plain}
\newfloat{sidefig}{thp}{lop}
\floatname{sidefig}{}

\newcommand{\degree}{^o}
\newcommand{\subsun}{_{\sun}}
\newcommand{\Break}{_{\mathrm{break}}}

\newcommand{\ICM}{_{\mathrm{ICM}}}
\newcommand{\Ram}{_{\mathrm{ram}}}

\newcommand{\DM}{_{\mathrm{DM}}}
\newcommand{\KeV}{\,\textrm{keV}}
\newcommand{\PC}{\,\textrm{pc}}
\newcommand{\Kpc}{\,\textrm{kpc}}
\newcommand{\Mpc}{\,\mathrm{Mpc}}
\newcommand{\Kms}{\,\textrm{km}\,\textrm{s}^{-1}}
\newcommand{\K}{\,\mathrm{K}}
\newcommand{\ccm}{\,\mathrm{cm}^{-3}}
\newcommand{\gccm}{\,\mathrm{g}\,\mathrm{cm}^{-3}}

\newcommand{\Myr}{\,\mathrm{Myr}}
\newcommand{\Gyr}{\,\mathrm{Gyr}}
\newcommand{\Reyn}{\textrm{Re}}

\shorttitle{Stripping of M89}
\shortauthors{Roediger et al.}

\begin{document}


\title{Stripped elliptical galaxies as probes of ICM physics: I. Tails, wakes, and flow patterns in and around stripped ellipticals}


\author{E.~Roediger\altaffilmark{1,2,5},  R.~P.~Kraft\altaffilmark{2}, P.~E.~J.~Nulsen\altaffilmark{2}, W.~R.~Forman\altaffilmark{2}, M. Machacek\altaffilmark{2}, S. Randall\altaffilmark{2}, C.~Jones\altaffilmark{2},  E.~Churazov\altaffilmark{3}, R.~Kokotanekova\altaffilmark{4}}
\affil{
\altaffilmark{1}Hamburger Sternwarte, Universit\"{a}t Hamburg, Gojensbergsweg 112, D-21029 Hamburg, Germany\\
\altaffilmark{2}Harvard/Smithsonian Center for Astrophysics, 60 Garden Street MS-4, Cambridge, MA 02138, USA\\
\altaffilmark{3}MPI f\"{u}r Astrophysik, Karl-Schwarzschild-Str. 1, Garching 85741, Germany\\
\altaffilmark{4}AstroMundus Master Programme, University of Innsbruck, Technikerstr. 25/8, 6020 Innsbruck, Austria
}

\email{eroediger@hs.uni-hamburg.de}


\altaffiltext{5}{Visiting Scientist, SAO}

\begin{abstract}
Elliptical cluster galaxies are progressively stripped of their  atmospheres due to their motion through the intra-cluster medium (ICM). Deep X-ray observations reveal the fine-structure of the galaxy's remnant atmosphere and its gas tail and wake. This fine-structure depends on dynamic conditions (galaxy potential, initial gas contents, orbit through the host cluster), orbital stage (early infall, pre-/post-pericenter passage),  and  ICM plasma properties (thermal conductivity, viscosity, magnetic field structure). We aim to disentangle dynamic and plasma effects in order to use  stripped ellipticals as probes of ICM plasma properties.  This first paper of a series investigates the hydrodynamics of progressive gas stripping by means of  inviscid hydrodynamical simulations. We distinguish a long-lasting initial relaxation phase and a quasi-steady stripping phase. During quasi-steady stripping, the ICM flow \textit{around} the remnant atmosphere resembles the flow around solid bodies, including a  `deadwater' region in the near wake.  Gas is stripped from the remnant atmosphere predominantly at its sides via Kelvin-Helmholtz instabilities. The downstream atmosphere is largely shielded from the ICM wind and thus  shaped into a tail. Observationally, both,  this `remnant tail' and the stripped gas in the wake can appear as a `tail', but only in the wake can galactic gas mix with the ambient ICM.    While the qualitative results are generic, the simulations presented here are tailored to the Virgo elliptical galaxy M89 (NGC 4552) for the most direct comparison to observations. Papers II and III of this series describe the effect of viscosity and compare to Chandra and XMM-Newton observations, respectively.
\end{abstract}


\section{Introduction} \label{sec:intro}
Elliptical galaxies falling into clusters are {progressively} stripped of their gaseous atmospheres due to their motion through the intra-cluster medium (ICM) (\citealt{Nulsen1982,Gisler1976,Takeda1984,Stevens1999,Toniazzo2001,Acreman2003,McCarthy2008}, among others).  Stripped ellipticals can display four basic features: A galactic atmosphere truncated in size, a downstream tail of  stripped galactic gas, an upstream contact discontinuity between the galactic atmosphere and the hotter ICM, and generally a bow shock ahead of the galaxy due to their typically supersonic motion.  {Elliptical galaxies with truncated atmospheres are commonly found in clusters and groups (\citealt{Sun2007,Jeltema2008}). X-ray bright tails of stripped gas are less common, but have been observed for several nearby galaxies already with early X-ray missions} (e.g., M86 and M49 in Virgo, \citealt{Forman1979}, \citealt{Irwin1996}, respectively). The bow shock is generally too faint { to be observed}, even if favorably oriented. 

Chandra and XMM-Newton observations have now reached a quality to reveal the details of the gas removal process and a wealth of structures {for several nearby cluster ellipticals}: 
M89 in the Virgo cluster has a curved tail and two `horns' attached to its upstream edge, bending downstream (\citealt{Machacek2006a}).
M86 in Virgo shows a spectacular 150 kpc long bifurcating tail starting in a plume (\citealt{Randall2008}), M49 in Virgo (\citealt{Kraft2011}) has a ragged upstream edge and a flaring tail, and the tail of NGC 1404 in Fornax appears rather faint despite its projected proximity to the cluster center (\citealt{Machacek2005}). The  details and differences of the gas-stripped cluster ellipticals intriguingly suggest the opportunity to determine the still ill-constrained ICM transport properties (thermal conductivity and viscosity) and the structure of magnetic fields in the ICM because these ICM plasma properties shape the flows in and around the stripped galaxies. 

The patterns of viscous and inviscid flows around  solid bodies are well-known. Recall, e.g., the flow  past a circular cylinder or sphere (\citealt{vanDyke1982}, among others): 
From high to low viscosity, or from low to high Reynolds number%
\footnote{The Reynolds number is defined as $\Reyn=\frac{l v \rho}{\mu}$ where $l$ and $v$ are the characteristic length scale and velocity of the flow, and $\rho$ and $\mu$ are the fluid density and dynamic viscosity.}%
, flow patterns change from laminar to a downstream vortex pair or torus to a turbulent wake. 

The ICM flow past the atmosphere of an elliptical galaxy should behave similarly, and  the observed structure of a stripped atmosphere and its wake should trace these flow patterns. The most obvious feature should be the structure of the wake: In a turbulent wake the stripped gas should mix quickly with the ambient ICM, reducing the gas density and hence X-ray brightness. {On the other hand, }a sufficient viscosity or aligned magnetic field suppresses Kelvin-Helmholtz instabilities (KHIs, \citealt{Roediger2013khi}, \citealt{Chandrasekhar1961}), which are a major agent of turbulent mixing. Therefore, in a sufficiently viscous or magnetized ICM, we expect an  unmixed, cool, X-ray bright tail. 

However, atmospheres of elliptical galaxies are not solid bodies. Instead, the atmospheres are compressible, bound by the galactic potential, and decrease in size with time due to the ongoing stripping. The ICM flow past the galaxy is not steady but varies in velocity and density along the galaxy's orbit through the cluster. In this first paper of a series, we identify characteristic stages and flow patterns around and inside elliptical galaxies undergoing inviscid gas-stripping. We do so by simulating an elliptical, or rather spherical, galaxy during cluster crossing as described in Sect.~\ref{sec:sim_setup}.  To facilitate the most direct comparison to observations, the simulations presented here are tailored to the stripped Virgo elliptical M89 regarding gravitational potential, gas contents, and orbit as described in App.~\ref{sec:tailor_m89}. The qualitative results are, however, independent of the exact galaxy model. In Sect.~\ref{sec:flowpatterns} we describe flow patterns and their evolution,  relate flow patterns in galaxy stripping to flow patterns around solid bodies, and suggest a consistent terminology.  In Sect.~\ref{sec:discussion} we briefly compare our results to previous work and discuss implications. Section~\ref{sec:summary} summarizes our results. In paper II of this series (Roediger et al.) we investigate the impact of an isotropic ICM viscosity on the stripping process and present observable signatures of viscous and inviscid stripping.  A detailed analysis of  new M89 Chandra and archival XMM data is presented in paper III (Kraft et al.), along with a comparison to the viscous and inviscid simulations.

 %
%
%
\section{Method}  \label{sec:sim_setup}

\subsection{The target galaxy M89}
%
\begin{figure}
\includegraphics[trim= 0 0 0 0,clip,angle=0,width=0.48\textwidth]{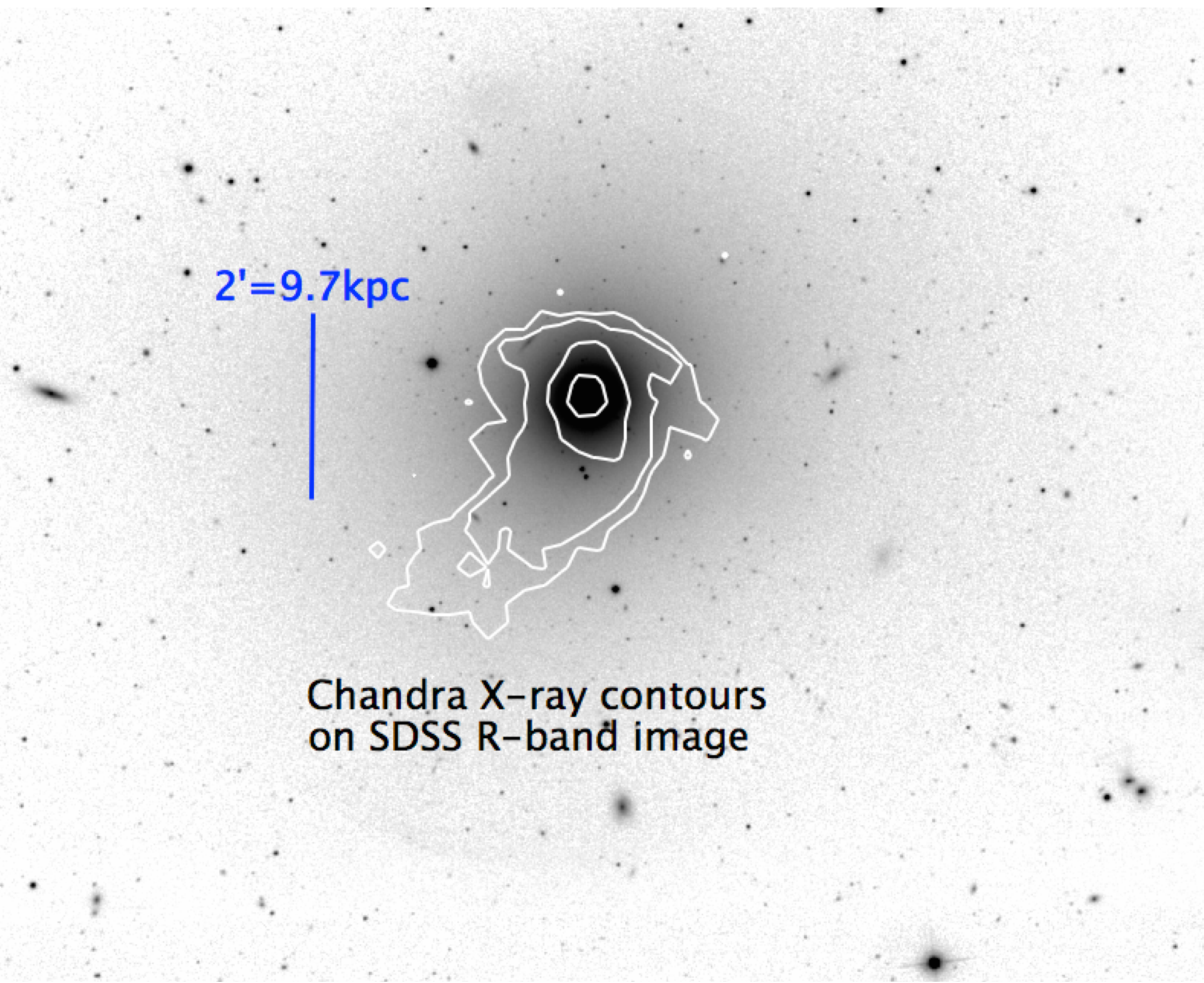}
\caption{Virgo elliptical M89. Chandra X-ray contours on SDSS R-band image (taken from GOLDMine, \citealt{Gavazzi2003}). Towards the north, the hot atmosphere is truncated down to $\sim 4$ kpc, but has a $\gtrsim $10 kpc tail to the south.}
\label{fig:m89optxray}
\end{figure}

{M89 is a massive elliptical galaxy located 350 kpc (72 arcmin) east of the Virgo cluster center (M87).   M89's hot atmosphere is detected in X-ray emission and shows clear signs of ongoing gas stripping (Fig.~\ref{fig:m89optxray}). The atmosphere is truncated down to 4 kpc in the north and has  a $\gtrsim $10 kpc gas tail in the south, indicating a motion of M89 through the Virgo ICM towards the north.}

Its stellar light is slightly elongated along the NW-SE axis, but for simplicity we use a spherical galaxy model. The impact of an elliptical potential is presented in a forthcoming paper; it does not affect the general flow dynamics described below.  {As detailed in Appendix~\ref{sec:tailor_m89}, we use observations of the stellar light from 0.1 to 30 kpc (\citealt{Kormendy2009}) to determine the inner gravitational potential of the galaxy, and estimate the total gravitating mass from the stellar velocity dispersion, the observed gas temperature in M89, and by comparing to field elliptical galaxies. We set an initial galactic gas temperature of 0.4 keV, close to the observed value (see paper III). The initial galactic gas density profile is matched to the observed values between 1 and 3 kpc. M89's outer atmosphere is already stripped, and its initial extent cannot be determined easily. Hence we simulate stripping of an initially compact and  an extended outer atmosphere to investigate the impact of this initial condition. M89's orbit through the Virgo cluster is estimated from its current position in the cluster, its radial velocity of $\sim 800\Kms$ with respect to the Virgo cluster, and its total velocity of $\sim 1700\Kms$ as measured from the stagnation point analysis (\citealt{Machacek2005}). Details are given in Appendix~\ref{sec:tailor_m89}.}

\subsection{Simulation setup}
We model the motion of a spherical galaxy -- M89 -- through its host cluster with 3D hydrodynamic simulations. The simulations are run in the rest frame of the galaxy, i.e., M89 is exposed to an ICM head wind which varies in density and velocity according to the galaxy's orbit though the Virgo cluster.  {We neglect gradients in the Virgo ICM and potential perpendicular to M89's orbit, {i.e., we model M89's motion through the Virgo cluster as a straight wind tunnel with a varying ICM head wind.} We treat M89's gravitational potential as static. At a pericenter distance of $\ge 350\Kpc$, M89's tidal stripping radius due to the Virgo cluster potential is $\sim 100\Kpc$, much larger than the size of the remnant atmosphere. Thus, tidal effects due to the cluster potential are unimportant for M89. Tidal effects may play a role for galaxies plunging deeper into their host potentials.}

The model galaxy consists of an analytic gravitational potential due to its dark matter and stellar contents, and an initially hydrostatic $\sim 0.4\KeV$ hot  atmosphere.  Tailoring the model to M89 is described in App.~\ref{sec:tailor_m89}. However, we tested a variety of galaxy models and found the flow patterns described below to be generic.

{Stellar mass loss in elliptical galaxies is expected to replenish the galactic atmosphere. \citet{Mathews1989} give a rate of $5.4\times 10^{-20}$ s$^{-1} \rho_*$ for an old stellar population, where $\rho_*$ is the spatial stellar density distribution. This corresponds to a total replenishment rate of $3.4\times 10^8 M_{\Sun}$ per Gyr for our M89 model. If replenishment could proceed undisturbed for 1 Gyr, the replenished gas mass would exceed the pre-existing one only in the central 4 kpc of M89, i.e., inside its current stripping radius. Hence the initially pre-existing atmosphere dominates the stripping features during first infall, and we neglect replenishment by stellar mass loss. \citet{Lu2011} study a replenishment-only case in a constant ICM wind.}

We model both the galactic atmosphere and the ICM as inviscid gases with an ideal equation of state. Thermal conduction and viscosity are neglected. The effect of viscosity is investigated in paper II.  We do not include radiative cooling or AGN heating of the galactic gas but assume that thermal balance is maintained by the interplay between both.

\subsection{Code}

We use the FLASH code (version 4.0.1, \citealt{Dubey2009}),   a modular, block-structured AMR code, parallelized using the Message Passing Interface library. It solves the Riemann problem on a Cartesian grid using the Piecewise-Parabolic Method. 

The size of our simulation box is $(-400\Kpc,400\Kpc)^3$, sufficiently large to capture the impact of the galaxy's potential on the ambient ICM. The galaxy is centered in the simulation box. The ICM wind enters along the $x$-axis.  The upstream boundary is an inflow boundary, where  ICM  density and temperature vary according to the current position of the upstream boundary in the cluster. The inflow velocity equals the opposite of the galaxy's current orbital velocity. The downstream boundary is an open boundary except for the outflow velocity being set equal to the inflow velocity. This ensures a correct ICM wind also in the subsonic regime. The side boundaries are open. ICM loss through them is negligible due to their large distance to the galaxy. 

FLASH offers adaptive mesh refinement on various conditions. We wish to resolve the galactic atmosphere and its removal and subsequent mixing with the ICM in the wake. As turbulence tends to be volume-filling, any useful refinement criteria would lead to refining  the complete wake region. We therefore use a manually adapted nested refinement around the galaxy and its tail, with a peak resolution of  0.1 kpc. The  peak resolution is ensured in the box of at least $(-10\Kpc, 22\Kpc)\times (\pm 10\Kpc)^2$ around the galaxy center. {This choice is motivated by the anticipated extent of the remnant atmosphere at pericenter passage.} {Only at late times} is the upstream extent of the peak resolution box  reduced to $\sim 2\Kpc$ ahead of the remnant atmosphere {to avoid unnecessary refinement of the free-streaming upstream ICM.} {The volume resolved to 0.2 kpc always encompasses the whole atmosphere and the  region downstream of the galaxy out to at least $\sim 45\Kpc$ from the galaxy center. We thus ensured that at pericenter passage even the  deadwater region for the initially extended atmosphere is refined to 0.2 kpc. } We confirmed that the resolution is sufficient for convergence of our results (see Fig. 16 in paper II).

\subsection{Seeding instabilities}

Instabilities grow from perturbations. We do not rely on seeding the instabilities by the numerical discretization, but introduce intentional seeds by repeatedly adding small perturbations to the ICM wind velocity  in the $x$-direction. The perturbation is a uniform superposition of sinusoidal waves of wave lengths 32, 16, 8, \ldots, 1 kpc, covering the size range from the initial atmosphere down to below the final stripped atmosphere. We chose a total amplitude of $50\Kms$ (few percent of sound speed) per perturbation event.  These perturbations are added every 300 Myr as they move out of the simulation box with the ICM wind.  While our perturbations are highly idealized in their power spectrum and direction, they serve the purpose of  seeds for the KHI and mixing. Similar to \citet{Shin2013}, we find that including perturbations enhances the gas removal rate. At the level of perturbations used here it is a secondary effect.

\subsection{Simulation runs}
Galaxy models and orbits are described in detail in App.~\ref{sec:tailor_m89}. Here we discuss the following simulation runs, sampling the uncertainty of M89's initial atmosphere and  orbital velocity: 
\begin{itemize}
\item galaxy with compact initial gas atmosphere ($n\propto r^{-1.5}$) on the slow orbit (pericenter velocity $1050\Kms=$Mach 1.3), 
\item galaxy with extended initial atmosphere ($n\propto r^{-1.2}$) on the fast orbit (pericenter velocity $1350\Kms=$Mach 1.7). 
\end{itemize}
The inner atmosphere for both galaxy models agrees with the one observed in M89.  In all cases the pericenter distance of the orbit is 350 kpc, equal to M89's projected distance to the Virgo center, which is very likely similar to the true cluster-centric distance as argued in paper III (see also App.~\ref{sec:ICMwind_obs}).  

The simulations start at a generous spatial and temporal distance of $\ge 0.8\Mpc$ and $\ge 1 \Gyr$ prior to the galaxy's pericenter passage,  and always with a subsonic initial ICM flow. This ensures that the stripping can reach a quasi-equilibrium state.

\section{Flow patterns and their evolution} 
\label{sec:flowpatterns}
%
{The goal of this paper is to determine  characteristic  flow patterns in and around gas stripped galaxies as well as their evolution.  We  first review classic concepts of gas stripping and show their limits (Sect.~\ref{sec:classic}). Second, we introduce the fluid dynamics approach to galactic gas stripping. To this end, we summarize flows past solid blunt bodies, focussing on characteristic flow patterns and flow stages (Sect.~\ref{sec:hydro_bluntbody}). Building on this knowledge, we define a corresponding terminology for the flow past a gas-stripped galaxy, and explain similarities and differences to the solid body case.   After having described the ICM flow \textit{around} the galactic  atmosphere, we will describe the evolution -- the stripping -- of the atmosphere itself. Finally, we describe the differences between stripping of an initially compact or initially extended gaseous atmosphere.}

\begin{figure*}
\includegraphics[angle=0,width=0.93\textwidth]{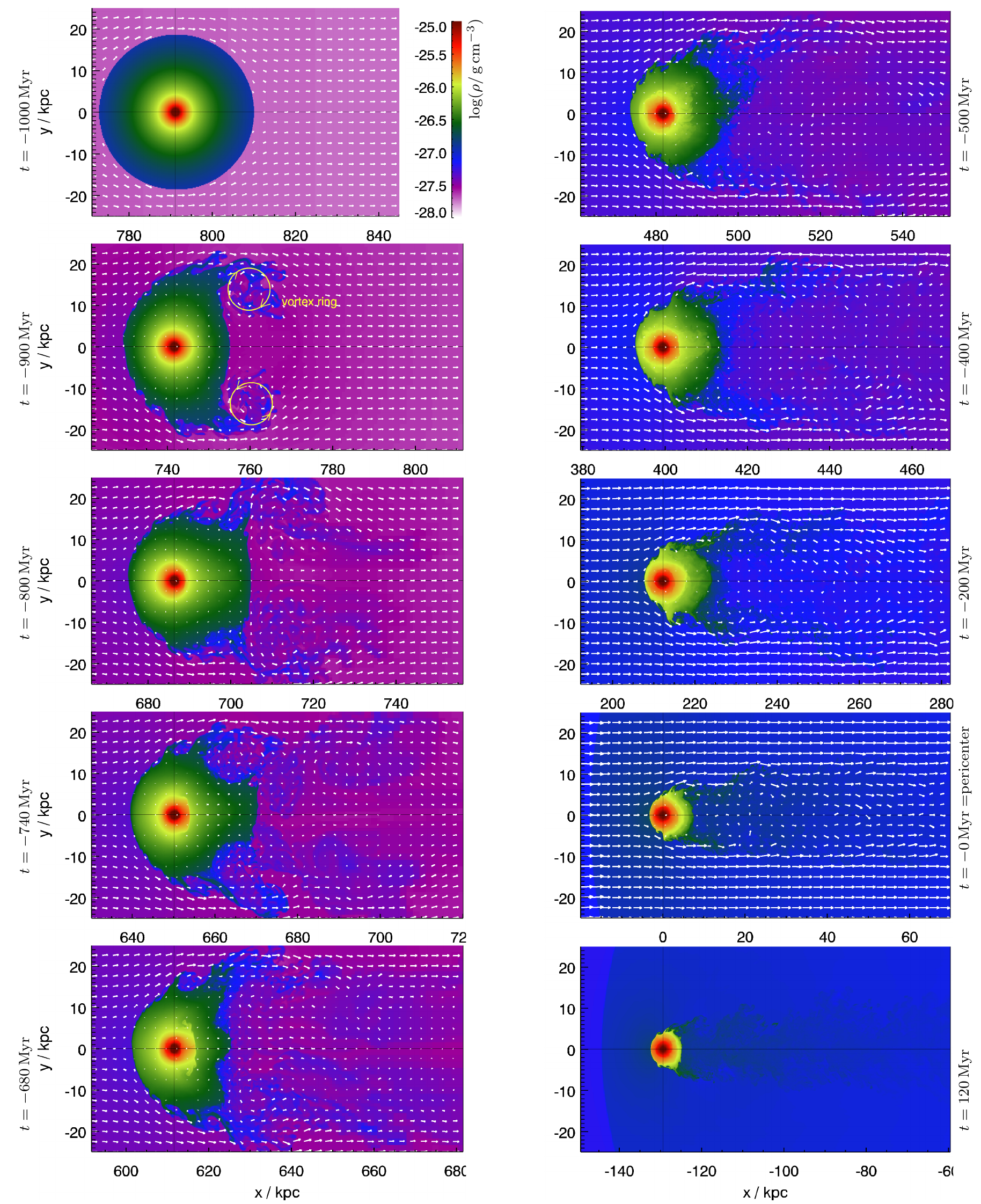}
%
\caption{Inviscid stripping of a spherical galaxy with an initially compact atmosphere. Density slices through the galaxy in the orbital plane, overlaid with velocity vectors (except last panel). The $x$-axis denotes the distance to  pericenter.   The left column shows the initial relaxation phase (Sect.~\ref{sec:flowpatterns}).  After the onset of the ambient flow, a vortex ring forms as marked in the second panel, and the downstream atmosphere oscillates back and forth twice. In the quasi-equilibrium phase (right column),  the flow patterns remain constant qualitatively, but their spatial scale decreases due to the ongoing stripping. The downstream atmosphere is shielded from the ICM flow and can be retained up to pericenter passage.  
Fig.~\ref{fig:vanDykeonly} contains a snapshot with flow structures labelled. Throughout the evolution, the upstream edge is pushed back, and KHIs remove gas from the sides of the galactic atmosphere.   Fig.~\ref{fig:radius_compact} shows the evolution of the atmosphere's radius towards the upstream, downstream, and side direction. 
\label{fig:dens_compact}}
\end{figure*}

\begin{figure}
\textbf{flow around a solid sphere:}
\newline
\phantom{x}\hfill\includegraphics[trim= 60 190 100 100,clip,angle=0,width=0.3\textwidth]{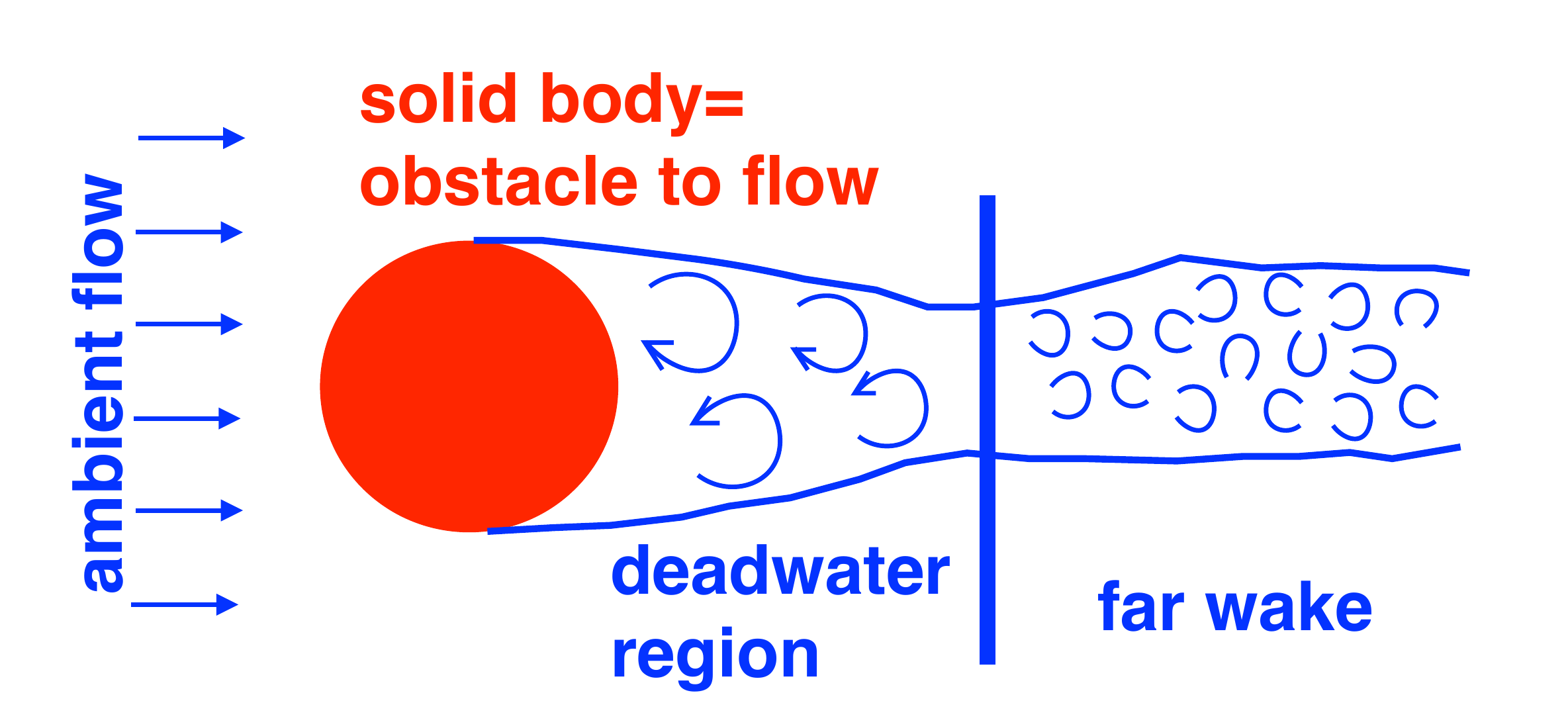}\hfill\hfill\phantom{x}
\newline
\newline
\textbf{flow around a stripped spherical galaxy, density of initial atmosphere $\propto r^{-1.5}$  (compact):}
\newline
\includegraphics[trim= 230 50 700 190,clip,angle=0,width=0.4\textwidth]{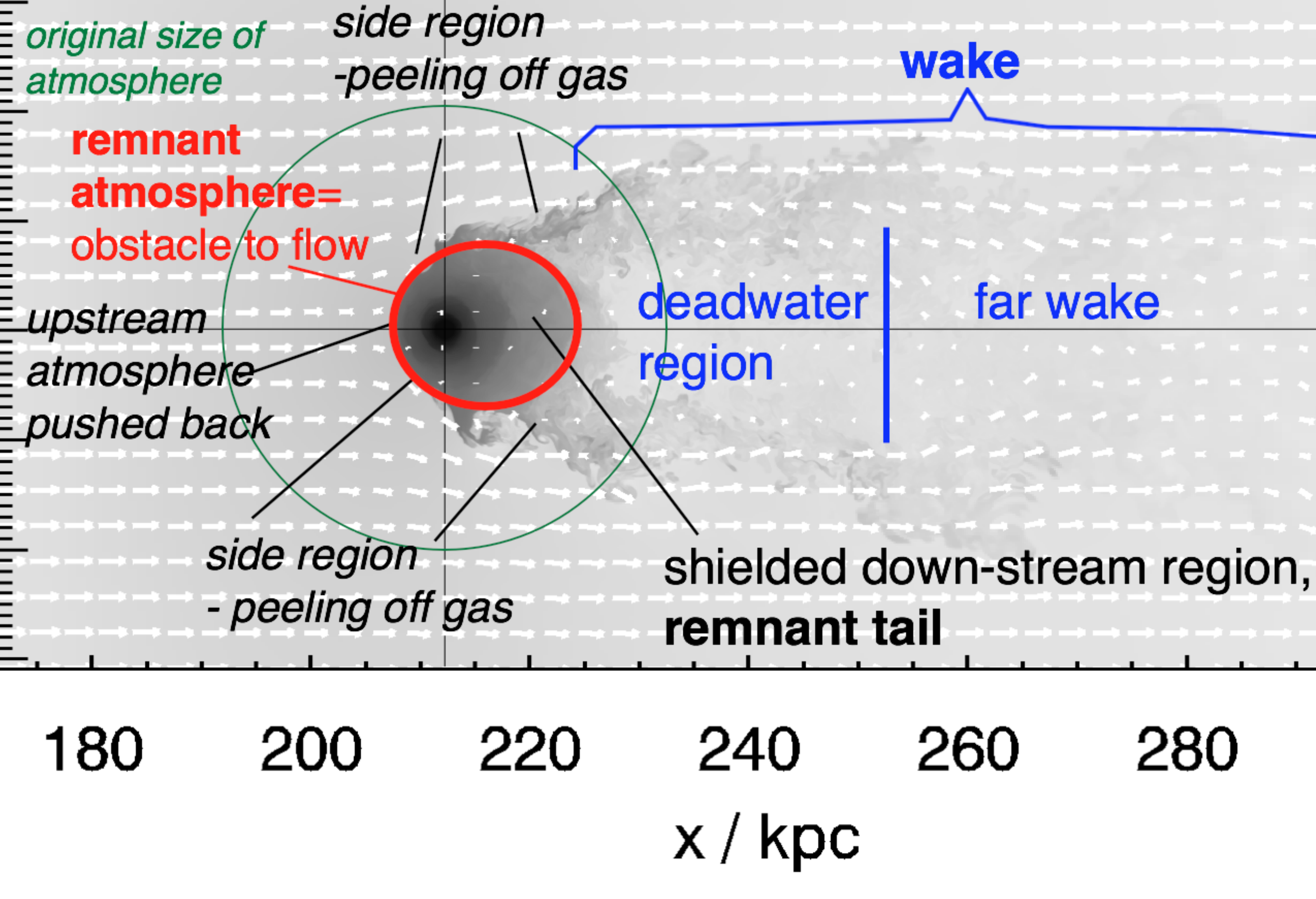}
\newline
\textbf{flow around a stripped spherical galaxy, density of initial atmosphere $\propto r^{-1.2}$ (extended):}
\newline
\includegraphics[trim= 280 40 570 150,clip,angle=0,width=0.44\textwidth]{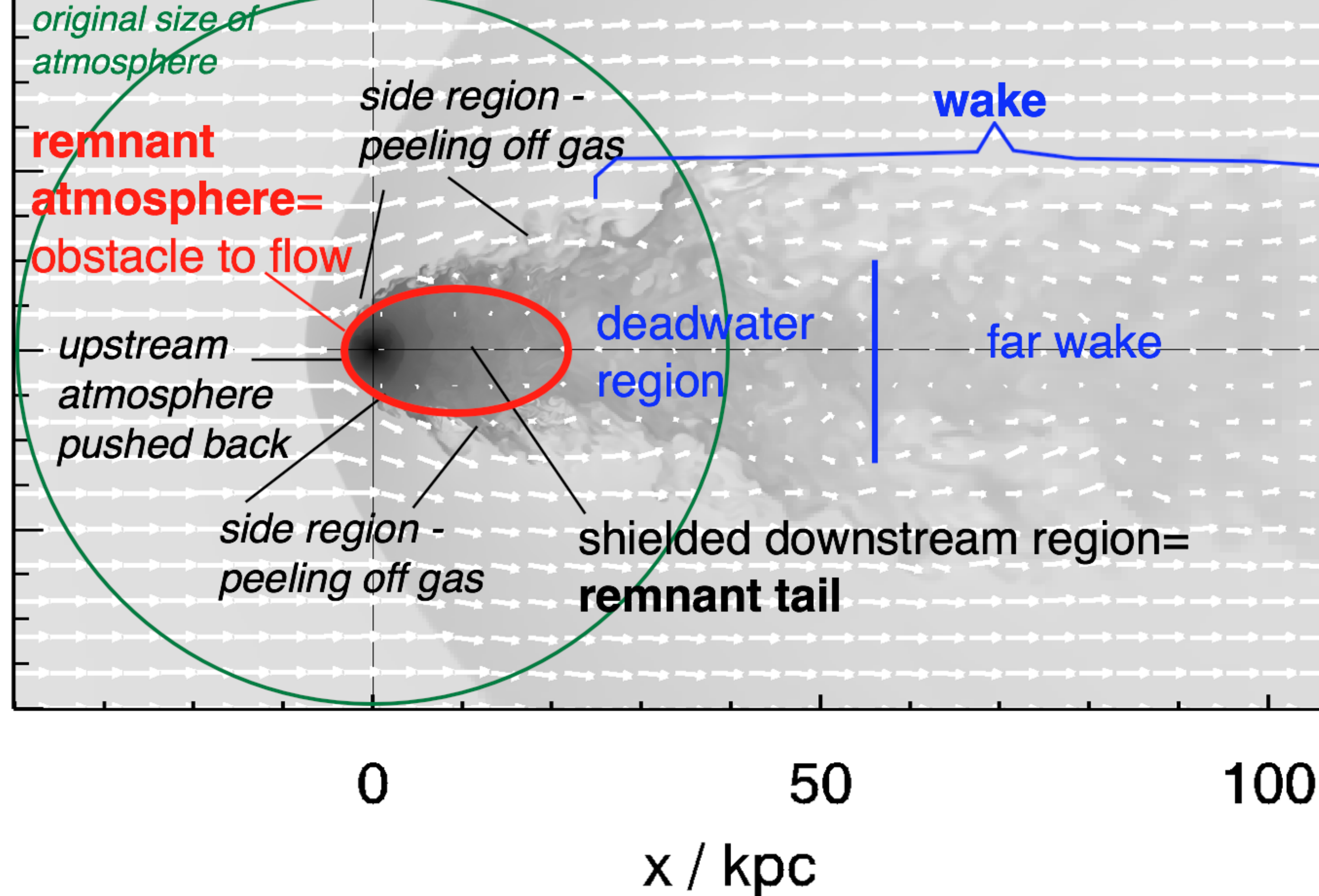}
\caption{Comparison of flow patterns around a solid body and in elliptical galaxy stripping.  We label flow regions in the ambient gas in blue and in the galactic atmosphere in black, see Sect.~\ref{sec:flowpatterns}. 
\newline
Top: {Sketch of a flow around a solid sphere, compare to, e.g.,  the photo of  the flow around a solid sphere in air in  Fig.~266 in \citet{vanDyke1982} (the sketch omits the bow shock seen in the photo).  Flow regions are labelled.} \newline
 Middle and bottom:  Slice through an inviscidly stripped spherical galaxy, gas density shown in grey-scale, velocity vectors overlaid. The green circle marks the initial extent of the galaxy's atmosphere. The center of the galactic potential is marked with the crosshair.   The middle panel is for the initially compact atmosphere at  200 Myr (212 kpc) prior to pericenter passage, moving with Mach 1.1; the bow shock just starts forming at $x\approx 180\Kpc$ and is still very faint. The bottom panel is for the initially extended atmosphere at pericenter passage, moving with Mach 1.7; the bow shock is clearly visible.  \newline
Qualitatively, flow patterns {around} the solid sphere and \textit{around} the stripped atmospheres are very similar. The stripped atmospheres retain a remnant tail that is part of the remnant atmosphere.  The extent of the current atmosphere is marked by the red ellipse. The whole atmosphere including the remnant tail poses the obstacle to the ambient ICM flow as does the solid body in the top panel.    The remnant tail is longer for an initially extended atmosphere. The longer remnant tail leads to a longer deadwater region as well. \newline
\label{fig:vanDykeonly}}
\end{figure}

\begin{figure*}
\begin{sideways}
\begin{minipage}{\textheight}
\includegraphics[angle=0,width=0.99\textwidth]{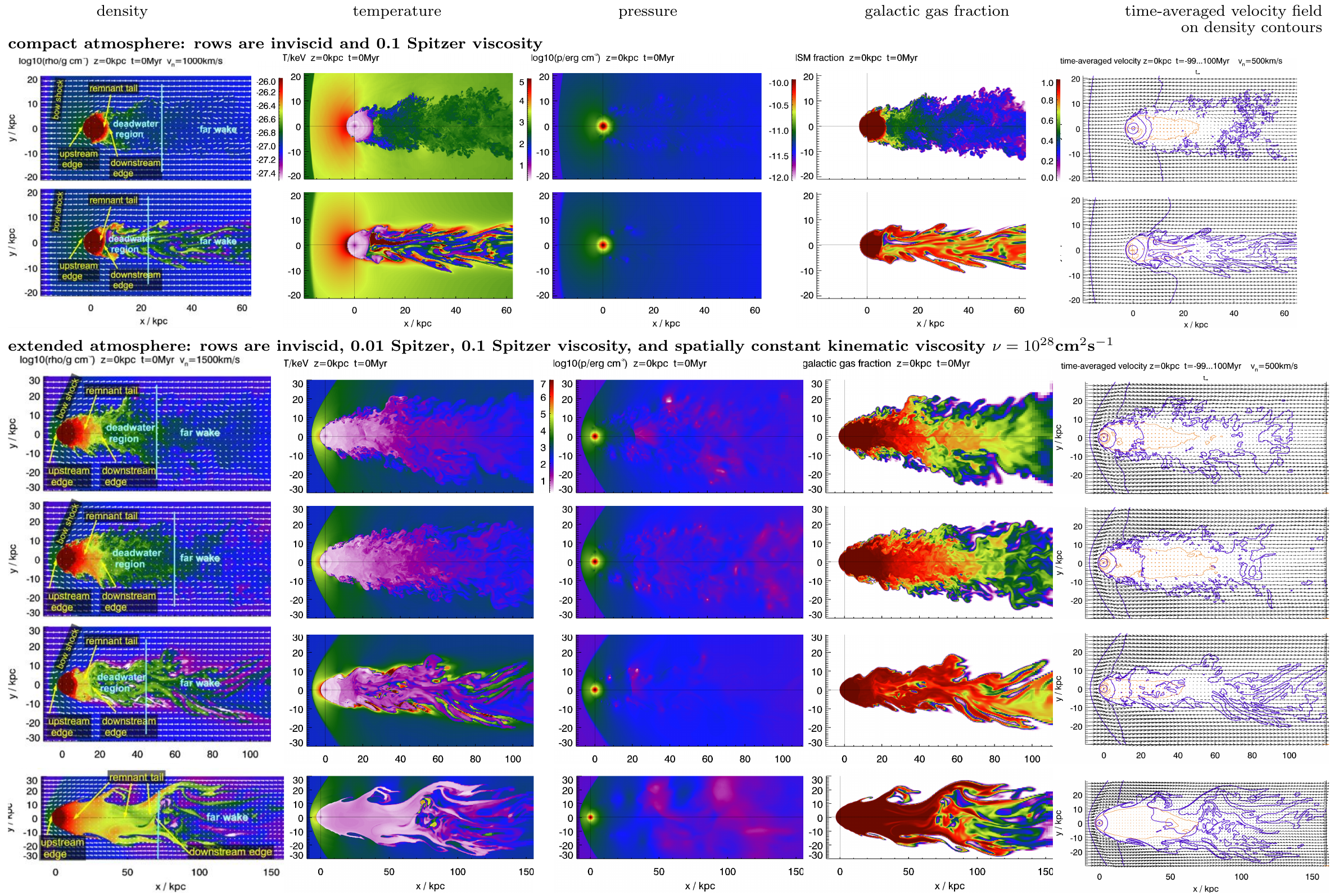}
\caption{Comparison of flow patterns of a stripped spherical galaxy; type of galaxy atmosphere and ICM viscosity as labelled.  The panels show slices through the 3D simulation box, color-coding the gas density, temperature, pressure, {and galactic gas fraction from left to right. The 5th column shows the time-averaged velocity field on density contours (blue), where upstream-directed velocities are plotted in red and are encompassed in the red contour line{ (deadwater region)}. The color scale for each quantity is shown in the first row and is valid for all following rows except for the temperature, where a new color bar is shown in the 3rd row.  The galactic center is at $(0,0)$ in all panels, and the stage at pericenter passage ($t=0$) is shown.} In the left column velocity vectors are overlaid and flow patterns labelled. Global flow patterns are the same in all cases, but the position and size of the deadwater region depends on the length of the remnant atmosphere. {Viscosity determines the mixing rate throughout the wake. The size of the panels differs between groups of rows to show the full flow patterns in each case.}  See Sect.~\ref{sec:flowpatterns} in this paper and Sect. 3 in paper II for details.
\label{fig:overview}}
\end{minipage}
\end{sideways}
\end{figure*}

{In the course of this section, we will refer repeatedly to three key figures.  
Figure~\ref{fig:dens_compact} sets the stage by showing evolution snapshots of gas stripping of the initially compact atmosphere.
  Figure~\ref{fig:vanDykeonly} compares the flow past a solid sphere to the flow past stripped galactic atmospheres. Figure~\ref{fig:overview} shows a side-by-side comparison of snapshots of selected gas properties from different simulation runs, varying dynamical conditions (here compact or extended atmosphere) and ICM properties (here viscosity).  Details of the viscous stripping are described in paper II.}


\subsection{Classic concepts of gas stripping} \label{sec:classic}
Earlier work on gas stripping described two mechanisms that remove gas from a galaxy placed in an ambient wind: the classic ram pressure stripping or pushing, and continuous stripping either due to KHIs or viscous momentum transport at the surface of the galactic atmosphere. 
\begin{itemize}
\item Classic ram pressure stripping was first invoked for the stripping of gas disks from disk galaxies.  It occurs in the outer regions of the galaxy, outside of the stripping radius, i.e., where the ram pressure of the ambient wind exceeds the local gravitational restoring force of the galactic potential on the galactic gas. This is the classic Gunn \& Gott criterion (\citealt{Gunn1972}). In this simple picture, the outer gas is virtually pushed out of the galaxy by the ambient wind because it is not sufficiently bound by gravity. Numerical simulations confirmed that this toy model can be used to estimate the stripping radius of disk galaxies (e.g., \citealt{Schulz2001,Roediger2005,Roediger2006}). The simulations also demonstrated that the ambient wind needs to continue pushing the outer gas for a few tens of Myr to accelerate the gas sufficiently to  escape  from the galaxy's gravitational potential.  \newline
If this classic ram pressure pushing would be the only process at work, no further gas loss would occur after the  gas outside the stripping radius reached escape velocity.
\item In the hydrodynamic description continuous stripping occurs. The shear flow between the ambient ICM wind and the galactic atmosphere (or gas disk) transfers momentum into the outer layers of the galactic atmosphere, thus progressively peeling off the outer layers from the atmosphere. The momentum transfer from the ICM into the galactic gas can occur via KHIs and via viscous momentum transfer, depending on the local viscosity. \citet{Nulsen1982} described these two varieties of continuous stripping as turbulent and viscous continuous stripping, respectively, and showed that 
the resulting gas loss rates are comparable. \newline
Continuous stripping occurs at all times and is also superimposed on the ram pressure pushing.
\end{itemize}
The term `ram pressure stripping' has been used inconsistently in the literature,  referring to either gas stripping in general, or referring specifically to the pushing mechanism described above. Neither of the two concepts above includes the flow of the ambient ICM around the remaining atmosphere.

Many simulations of gas stripping used a constant ICM wind flowing past a galaxy to focus on specific aspects of the stripping process. Given that the ram pressure pushing occurs on timescales of only a few 10 Myr, rather short compared to a galaxy's lifetime, gas-stripping can appear to be a two-stage process -- first, ram pressure pushing removes the outer disk (sometimes also called instantaneous stripping), then the galaxy enters the continuous stripping phase. 

While this two-stage concept captures the basics of gas stripping in a \textit{constant} wind, it is not correct. The onset of the ram pressure pushing indeed follows closely the toy model outlined above. However, in a hydrodynamic flow, some of the stripped gas can move into the region `behind' the remaining galactic atmosphere where it is shielded from the ICM wind. Depending on the details of the dynamics, such shielded gas can  fall back to the remaining atmosphere, or, if the gas is moving in and out of the wind shielding, the timescale to accelerate gas to escape velocity is extended to $\sim 100 \Myr$ or more. 
In addition, the continuous stripping mechanism described above is at work at all  times because the shear flow between the ICM and the galactic gas exists continuously. Only the lower gas loss rate of the continuous stripping leads to the appearance of continuous stripping as a separate, second stripping stage.

{For comparing with observations it is important to account for the fact that }galaxies falling into clusters do not experience a constant, but an intensifying, ICM head wind, i.e., the ICM head wind density and velocity increase. Consequently, stripping by an increasing ICM wind cannot be described by a single ram pressure pushing event. Instead, we have to expect a continuum of overlapping ram pressure pushing events as the ICM ram pressure increases. The continuous stripping mechanism will be superimposed at all times.

Figure~\ref{fig:dens_compact} shows the evolution of gas stripping of the initially compact atmosphere during the galaxy's infall into the Virgo cluster.  The figure covers the evolution from 1 Gyr prior to pericenter passage to 120 Myr after pericenter passage. We recover the basic expectations drawn from the classic concepts outlined above. The galactic atmosphere is stripped to a progressively smaller size  as the ICM flow around it gets denser and faster. Stripped-off gas is eventually carried away by the ICM flow. In this inviscid simulation, the continuous stripping mechanism operates via KHIs that are ubiquitous at the surface of the atmosphere along its side.  

However, certain aspects of the stripping are understood better with a fluid dynamics approach, e.g., the oscillations of the downstream atmosphere in the left column in Fig.~\ref{fig:dens_compact}, the fact that the downstream radius of the atmosphere  always exceeds the upstream radius, and the region of very low or even negative flow velocity downstream of the galactic atmosphere.

\subsection{Hydrodynamics approach to gas stripping -- the flow around the galactic atmosphere} \label{sec:hydro_bluntbody}

From the hydrodynamics perspective, gas-stripping of an elliptical or spherical galaxy can be compared to the flow around a blunt, solid body. At any given time, the remaining galactic atmosphere takes the role of the blunt body as it poses an obstacle to the ambient ICM flow just as a blunt body does.    However, we need to account for the fact that the galaxy's atmosphere is not a solid body but a gaseous atmosphere bound by gravity, and shrinking in size due to the gas stripping.  Despite these complexities, direct analogies to the well-known flows around solid bodies can be made.

\subsubsection{Quasi-steady flow past blunt body and past a stripped galaxy} \label{sec:hydro_steady}

{The top panel of Fig.~\ref{fig:vanDykeonly} shows a sketch of a quasi-steady flow past a solid sphere. The sketch follows the schlieren photo of  the flow around a solid sphere in air in  Fig.~266 of \citet{vanDyke1982}. Note that in the sketch we omitted the bow shock seen in the photo.} 
At the downstream side, the flow is characterized by a wake, which here is turbulent as in the shown case the  Reynolds number is high.

However, even in such  turbulent wakes, the \textit{average} flow follows a regular pattern. The average flow in the near wake is torus-like  (e.g., Fig.~56 in \citealt{vanDyke1982}). This toroidal flow is oriented such that the outer part of the torus flows away from the sphere, i.e., it is aligned with the global ambient flow. 
Consequently, in the inner part of the torus, or in the inner part of the near wake, the average flow velocity is directed \textit{upstream} in the rest frame of the sphere. The average flow velocity in this torus is low compared to the ambient flow.  Such `backflow' or `deadwater' regions are obvious in the wakes of cars or trucks -- they can drag along dust or dry leaves behind them. Similar flow patterns occur in the wakes of boats.  In the far wake the mean gas flow is again directed away from the moving body. 

Qualitatively, these \textit{average} wake flow patterns are independent of the shape of the blunt body and are also independent of Reynolds number except for laminar or creeping flows at very low Reynolds number below $\lesssim 20$ (\citealt{vanDyke1982,Taneda1956}) which lack the deadwater region. However, the shape of the blunt body, its surface properties,  and the Reynolds number of the flow influence the extent of the deadwater region. For solid spheres, the deadwater region can extend as far as one or more sphere diameters  downstream of the sphere (\citealt{Taneda1956}).  At moderate Reynolds numbers around 100, the deadwater region takes the shape of a vortex ring or vortex pair in 3D and 2D, respectively (e.g., \citealt{Taneda1956,Taneda1956b}). 
At Reynolds numbers above $\sim 1000$ both {the} deadwater region and far wake {are} superimposed by turbulence in the wake  {(e.g., Fig.~266 in \citealt{vanDyke1982})}.

\begin{figure}
\includegraphics[width=0.48\textwidth]{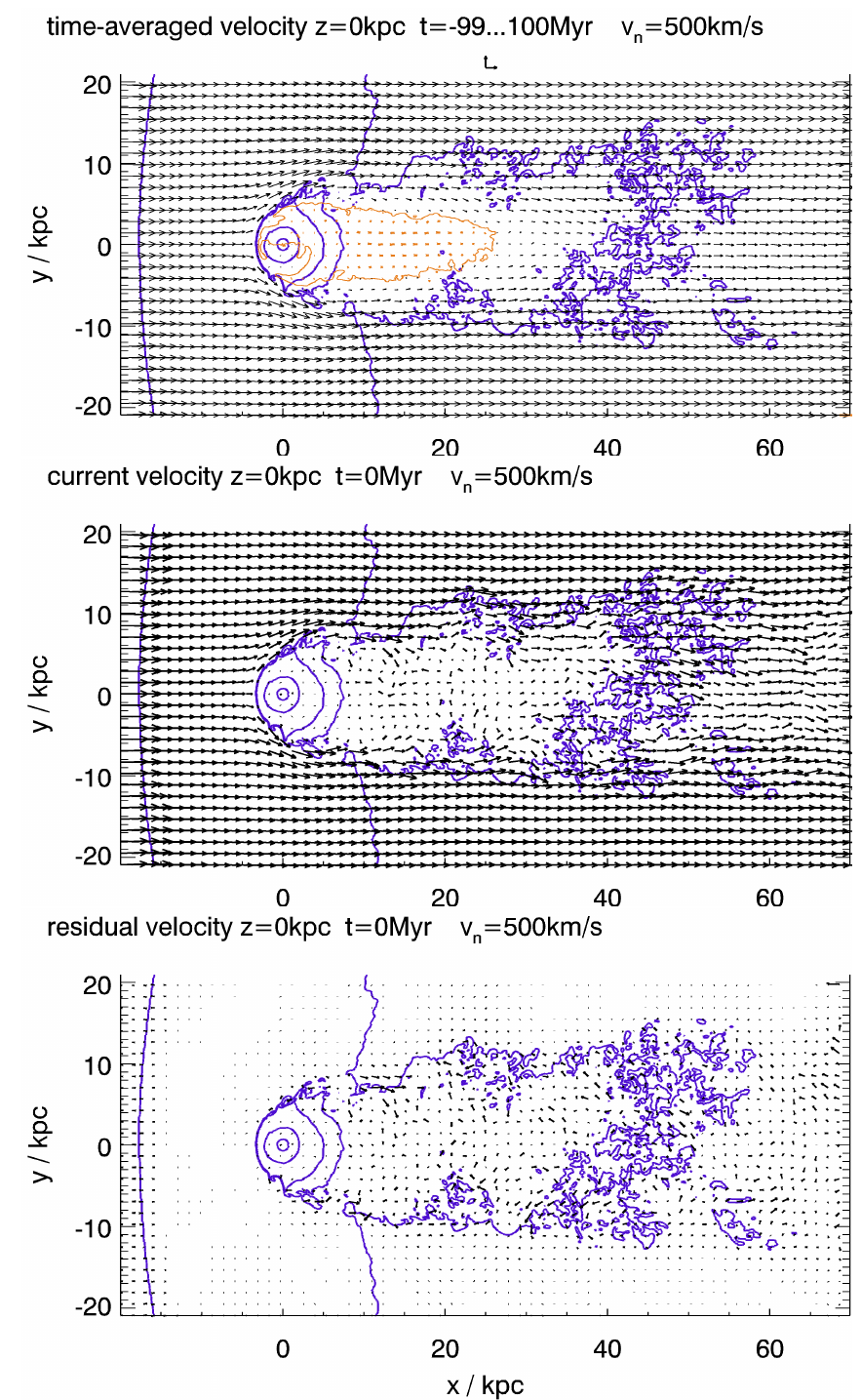}
\caption{Top: Averaged velocity field around a gas-stripped elliptical galaxy. The plot shows the initially compact atmosphere at pericenter passage, see second last panel in Fig.~\ref{fig:dens_compact}. Blue contours show the gas density (logarithmic spacing with factor $\sqrt{10}$ between densities $10^{-27}$ and $10^{-25}\gccm$). The velocity field is averaged over $\pm 100\Myr$ around pericenter passage. Upstream-directed velocity vectors are plotted in red, and the red contour encompasses regions with upstream-directed velocities. The region in the near wake with negative velocities is defined as the deadwater region. Middle: velocity field at pericenter passage on density contours. Bottom: residual velocities with respect to the averaged velocity field, showing the turbulence superimposed to the average flow. Residuals near the bow shock arise because the position of the bow shock moves slightly over the averaging interval.}
\label{fig:mediumflow}
\end{figure}
%

We can revisit the gas stripped galaxy shown in Fig.~\ref{fig:dens_compact} with 
 the flow patterns past a solid body in mind. In all panels in the right column, we can identify the galaxy's wake in the velocity field. The second panel in Fig.~\ref{fig:vanDykeonly} shows a labelled snapshot of this simulation in direct comparison to the solid sphere case.
 In the near wake just downstream of the galactic atmosphere, the flow velocity is very low compared to the ambient ICM flow. Averaging the velocity field over time reveals the torus-like average flow pattern in the near wake as shown in Fig.~\ref{fig:mediumflow}.  The red contour in the averaged velocity field encompasses regions with a negative velocity in $x$-direction. We define this region as the backflow or deadwater region, i.e., the region of the near wake with an, on average, upstream-directed velocity. Downstream of the deadwater region the far wake begins where the flow velocity is directed away from the galaxy. 
 The deadwater region is generally about 1.5 times as long as the total length of the remnant atmosphere, in qualitative agreement with laboratory results for solid spheres (\citealt{Taneda1956}). 
 
 The wake of a supersonic solid body is usually about as wide as the cross-section of the body. The wake of a stripped galaxy tends to be somewhat wider than the remnant atmosphere because the atmosphere itself is continuously shrinking.   The galaxy's wake is in pressure equilibrium with the ambient ICM except for scattered low-pressure regions inside vortices.

\subsubsection{Onset of the flow vs. quasi-steady flow} \label{sec:hydro_phases}

Section~\ref{sec:hydro_steady} above describes the quasi-steady flow past a solid body or a galactic atmosphere.
However, it is well-known from experiments that an impulsive onset of a flow is not immediately followed by a quasi-steady flow, but by a relaxation phase. For an impulsive onset of a flow around a solid body, the relaxation phase lasts several object crossing times (=length/velocity) before the flow  settles into  a quasi-steady state. The relaxation phase proceeds in a characteristic series of stages: Right after the onset the flow appears laminar. Next, a downstream vortex ring forms. Then the wake starts shedding vortices or turns turbulent for sufficiently large Reynolds numbers (see Fig. 59 in \citealt{vanDyke1982}, or Fig.~5.11.3 in \citealt{BatchelorHydro} (taken from \citealt{Prandtl1927})). Finally, a quasi-steady wake is established.

During a galaxy's cluster passage, the ICM head wind varies in density and velocity by a factor of several in both quantities, the ram pressure in the simulated part of M89's orbit varies by a factor of $\sim 50$. Due to the progressive stripping, the size of the galactic atmosphere shrinks dramatically.  As both ambient wind and  size of the `blunt body' in the flow vary, we expect -- and find -- an extended relaxation phase for galaxy stripping. 

\begin{figure}
\includegraphics[trim= 0 0 0 0,clip,angle=0,width=0.45\textwidth]{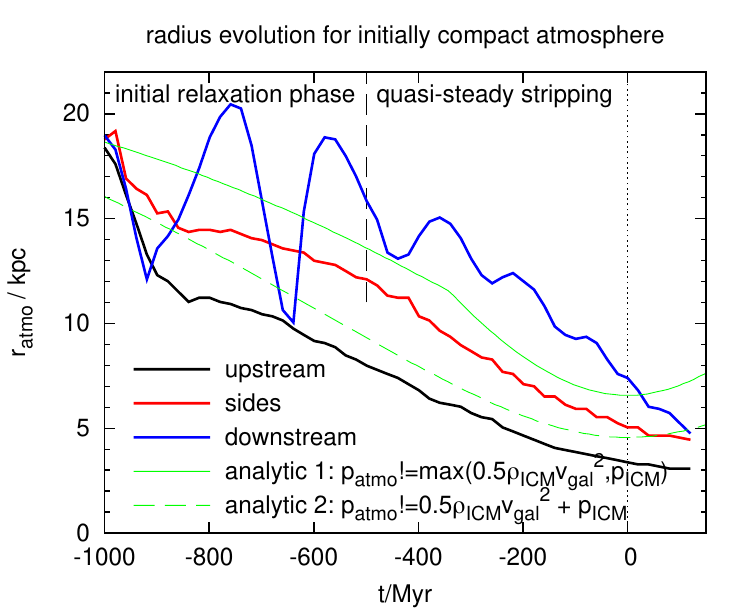}
\includegraphics[trim= 0 0 0 15,clip,angle=0,width=0.45\textwidth]{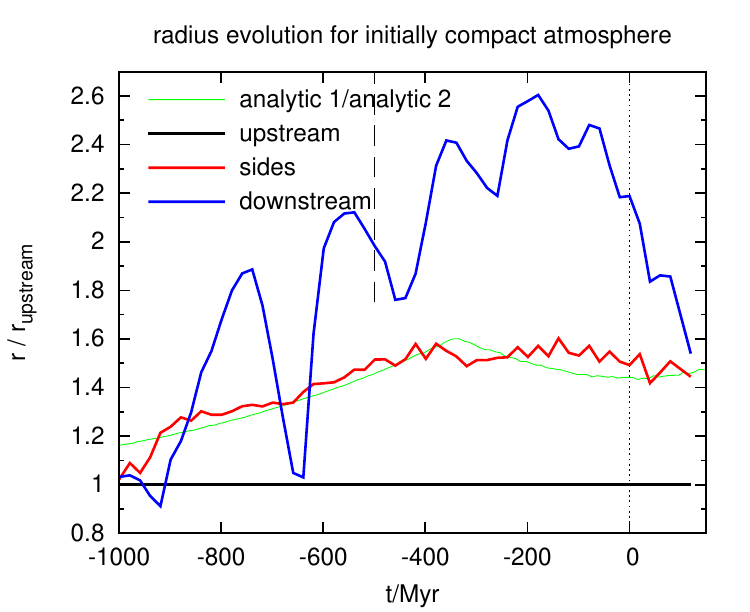}
\caption{Evolution of the size of the remnant atmosphere measured
 towards the upstream and downstream direction and towards the sides for the initially compact atmosphere (see Fig.~\ref{fig:dens_compact}). The top panel plots the radii in these three direction in kpc, the bottom panel is normalized to the upstream radius. In green, two versions of an analytic estimate of the stripping radius are shown: The stripping radius is estimated where the atmosphere's pressure $p_{\mathrm{atmo}}$ equals the maximum of ICM ram pressure, $0.5\rho\ICM v\ICM^2$, and thermal pressure, $p\ICM$, or their sum. These analytic estimates closely follow the trends for the side radius and upstream radius, but slightly over-predict them, {and their ratio closely describes the ratio of side radius to upstream radius as shown in the bottom panel.} The downstream radius first experiences strong oscillations, then remains significantly larger than the upstream and side radii until pericenter passage. {The vertical dashed line marks the transition from the initial relaxation phase to the quasi-steady stripping phase.}
\label{fig:radius_compact}}
\end{figure}

The left  column in Fig.~\ref{fig:dens_compact} shows this relaxation phase in galaxy stripping. In the second panel we marked the initial vortex ring. In the subsequent panels, this first vortex and subsequent ones are shed, which leads to the strong oscillations of the downstream atmosphere along the flow direction. The downstream oscillations of the galaxy's atmosphere are clearly evident when plotting the downstream radius of the atmosphere as a function of time, as shown in Fig.~\ref{fig:radius_compact}. 

In contrast, the right column of Fig.~\ref{fig:dens_compact} shows the quasi-steady flow phase where the flow patterns in all panels are very similar except for their spatial scale. In this quasi-steady stripping stage, the downstream atmosphere can still oscillate slightly, but these oscillations are only a secondary effect compared to the continuous decrease of the overall size of the atmosphere due to the ongoing gas stripping.

\subsubsection{Anatomy of a gas stripped galaxy}
Sections~\ref{sec:hydro_steady} and \ref{sec:hydro_phases} discussed the analogy between the flow past a solid sphere and past a gas stripped galaxy.  Before continuing with describing the evolution of the galactic atmosphere itself, we summarize the terminology introduced above.

An elliptical galaxy falling into a cluster experiences gas stripping by an ICM head wind of increasing intensity. The progressive gas stripping continuously reduces the size of the galaxy's atmosphere. The flow past the remaining atmosphere first undergoes a relaxation phase before entering a quasi-steady phase (see Figs.~\ref{fig:dens_compact} and \ref{fig:radius_compact}): 
\begin{itemize}
\item After the onset of the flow, a series of two or three vortex rings form downstream of the atmosphere and detach from the atmosphere. The vortex shedding induces strong oscillations of the downstream galactic atmosphere. {We note that the} very first isotropic compression of the atmosphere is an artifact of the initial setup. While the galactic atmosphere is set up in hydrostatic equilibrium with the galactic potential, the ambient ICM is not. Consequently, the ambient ICM falls onto the galactic atmosphere, compressing it somewhat before the stripping sets in.
\item The galaxy enters a quasi-steady stage  $\sim 500\Myr$ prior to pericenter passage. In the quasi-steady phase, the average flow patterns around the remaining atmosphere are constant except for decreasing in spatial scale due to the shrinking galactic atmosphere.   The evolution of the downstream extent of the remaining atmosphere is dominated by the ongoing stripping and not by oscillations due to vortex shedding, although these can still lead to minor oscillations of the downstream atmosphere.
\end{itemize}

In  quasi-steady stripping the flow around the remaining atmosphere closely resembles the quasi-steady flow past a solid blunt body:
\begin{itemize}
\item The near wake of the galaxy contains a backflow or deadwater region that is defined by a  slow flow directed upstream. The upstream-directed flow region is the inner part of a torus-like flow that characterizes the near wake.
\item In the far wake, i.e., downstream of the deadwater region, the average flow is directed away from the galaxy.
\item At sufficiently high Reynolds number the average wake flow patterns are superimposed by turbulence (bottom panel of Fig.~\ref{fig:mediumflow}). 
\item If moving supersonically, both solid bodies and stripped galaxies cause an upstream bow shock.
\end{itemize}

Throughout the description above we referred to the leftover atmosphere at any given time as `remaining' or `remnant' atmosphere. Occasionally the term `remnant merger core' has been used for gas-stripped atmospheres of galaxy groups or galaxies. However, here the main driver in shaping the galactic atmosphere is gas stripping and not gravitational, merger-related effects, unless the galaxy passes close to the center of its host cluster. Hence we prefer the term `remaining' or `remnant atmosphere'. Furthermore, we will show below that the remnant atmosphere can have a pronounced elongated shape which is not well-described by the term `core'.

{We followed the origin of all gas in the simulation by a passive tracer that was set to 1 inside the initial atmosphere and to zero in the ICM. Thus, this quantity traces the galactic gas fraction, and grid cells with a galactic gas fraction of 1 contain purely galactic gas. We can define gas with a  galactic gas fraction of 1 that is gravitationally bound to the galaxy as the remnant atmosphere.  This is the most conservative choice and results in the smallest remnant atmosphere. Alternatively, we can use an entropy threshold of  $50$ keV cm$^2$, which is roughly a factor of two above the maximal entropy in the galactic gas at the start of the simulations.   
In the stripping of the initially compact atmosphere (Fig.~\ref{fig:dens_compact}), remnant atmosphere is easy to identify in either gas density, entropy, temperature, or galactic gas fraction (first row  in Fig.~\ref{fig:overview}), and the extent of the remnant atmosphere is insensitive to the exact  threshold applied. 
For the initially extended atmosphere, the downstream boundary between the remnant atmosphere and the deadwater region in the wake is less well-defined. We measure the extent of the remnant atmosphere towards the upstream, downstream, and side directions by taking entropy profiles in $20\degree$ cones/slices towards each direction {(these measurements are used in Figures \ref{fig:radius_compact} and \ref{fig:radius_ext})}. Gas with entropy below $50$ keV cm$^2$ is defined as galactic gas.}

The remnant atmosphere generally extends farther into the downstream direction than into the upstream direction, which can give the downstream atmosphere a tail-like appearance. We will refer to the extended downstream part of the remnant atmosphere as the remnant tail (see also Sect.~\ref{sec:atmosphere}). The remnant tail is a striking feature for the initially extended atmosphere (e.g., Figs.~\ref{fig:dens_extended}, \ref{fig:vanDykeonly}). Also the stripped initially compact atmosphere has a remnant tail although it is not well-pronounced.

The flow patterns defined here are labeled in Fig.~\ref{fig:overview} for all simulation runs.

\subsection{Evolution of the remnant atmosphere}  \label{sec:atmosphere}
The ambient ICM flow has two effects on the remnant atmosphere: it strips off gas, and it deforms the remnant atmosphere in characteristically different manners at the atmosphere's upstream part, its sides, and its downstream part.  Below we discuss these  regions separately.

\subsubsection{Evolution of upstream atmosphere}

The evolution of the upstream atmosphere is governed by the pressure balance between the galactic pressure and the ram pressure of the ICM flow. The ICM ram pressure pushes back the upstream atmosphere until pressure balance is reached.
The pressure in the galactic atmosphere increases towards the galactic center, thus, the ICM ram pressure cannot push back gas from the galactic center where the galactic pressure exceeds the ram pressure. Thus, the ICM ram pressure pushes  the upstream atmosphere back to the radius where the ICM stagnation point pressure is balanced by the thermal pressure in the atmosphere (see pressure slices in Fig.~\ref{fig:overview}, also \citealt{Mori2000}). Thus, the upstream stripping radius  of the galaxy at a given time can be estimated by balancing the galactic gas pressure, $p_{\mathrm{atmo}}$, and the current stagnation point pressure, which is the sum of the thermal ICM pressure and the ram pressure, $p\ICM + p\Ram=p\ICM+0.5\rho\ICM v\ICM^2$ (dashed green line in Figs.~\ref{fig:radius_compact} and \ref{fig:radius_ext}).   

 After the onset of the flow, this pressure balance at the upstream edge of the atmosphere is reached quickly, within 100 Myr (Figs.~\ref{fig:dens_compact}, \ref{fig:radius_compact}, also Figs.~\ref{fig:dens_extended} and \ref{fig:radius_ext} for the extended atmosphere). As both ICM thermal and ram pressure increase during the galaxy infall, the upstream radius of the remnant atmosphere continues to decrease slowly up to pericenter passage (Figs.~\ref{fig:radius_compact}, \ref{fig:radius_ext}). In all our simulations, the final upstream radius is similar to the observed value for M89, which is a simple consequence of tailoring the galactic potential, inner atmosphere and ICM wind to this galaxy. 

\subsubsection{Evolution of the sides of the atmosphere}

\begin{figure}
\includegraphics[width=0.45\textwidth]{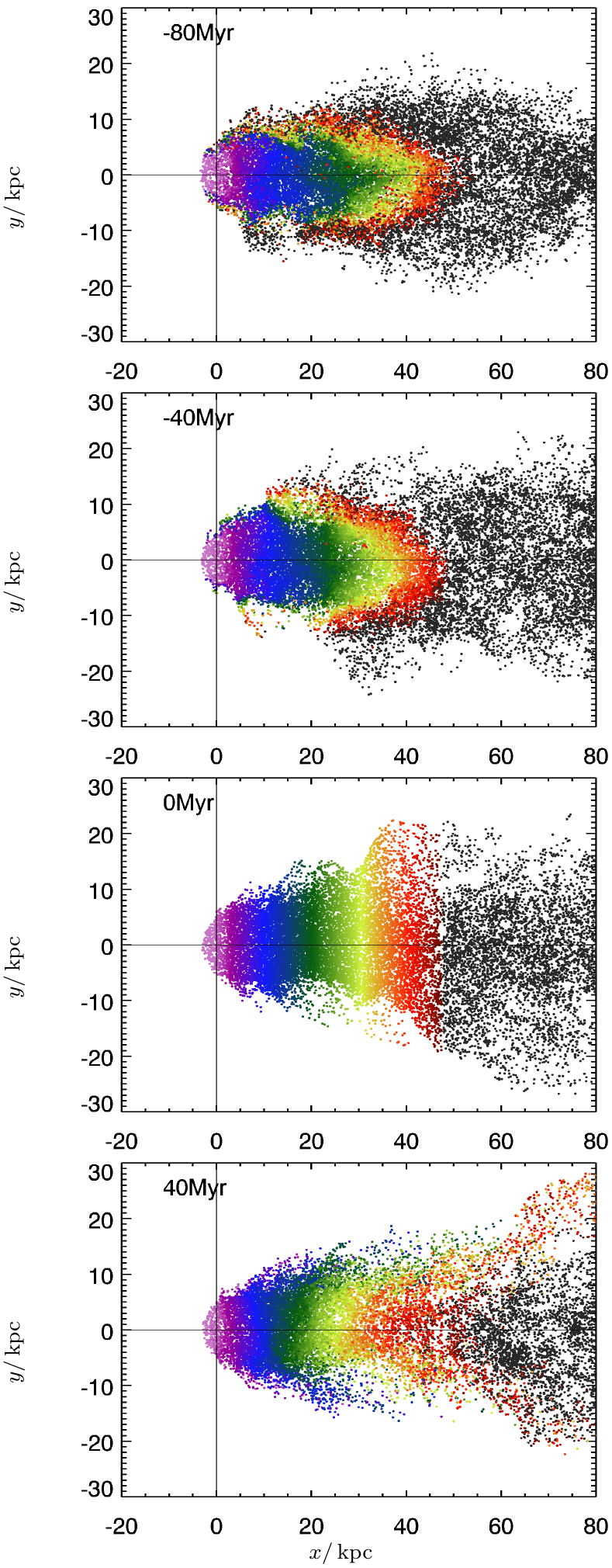}
\caption{Visualization of gas removal from galaxy. Axes are labeled in kpc. We selected tracer particles in the galaxy and wake in a  {6 kpc thick } slice around $z=0$ at pericenter passage ($t=0$) and color-coded them according to their position along $x$ at this time (3rd panel). The particles keep their color in previous and subsequent panels.  Gas removal from the galaxy occurs only at the interface between the galaxy and the ambient ICM, and the interface between the wake and the ICM, but not from the inner part of the wake or atmosphere.    This figure shows inviscid stripping of the initially extended atmosphere.
\label{fig:removal}}
\end{figure}

In gas stripping of the hot atmosphere of a $\sim$spherical galaxy studied here, the ram pressure pushing alone does not strip off galactic gas, but merely deforms the atmosphere.
The actual gas removal occurs along the sides of the galactic atmosphere where the shear flow, with respect to the ICM, is largest (Fig.~\ref{fig:dens_compact}).  In these shearing layers, turbulent or viscous momentum transport accelerate the outer layers of the atmosphere in the wind direction and  thus ablate gas from its sides.
In the inviscid simulations presented here,  KHIs transfer the momentum from the ICM wind into the atmosphere. Figure~\ref{fig:removal} visualizes this process using tracer particles. This continuous stripping mechanism is always present, it starts immediately after onset of the ICM wind and continues past pericenter passage (see also \citealt{Nulsen1982}).

%
%

The atmosphere's radius to the sides, i.e., perpendicular to the galaxy's direction of motion, is typically 1.5 times larger than the upstream radius (Figs.~\ref{fig:radius_compact}, \ref{fig:radius_ext}). 

\subsubsection{Evolution of the downstream atmosphere}

The downstream part of the atmosphere evolves differently during the initial relaxation phase and during  quasi-steady stripping. As indicated above, the vortex shedding during the initial relaxation phase leads to strong oscillations of the downstream extent of the atmosphere (Figs.~\ref{fig:dens_compact}, \ref{fig:radius_compact}). These oscillations arise because the atmosphere is compressible and adapts to alternating downstream pressure decrements and backflows during vortex shedding.
At about 500 Myr prior to pericenter passage, the flow around the remnant atmosphere reaches a quasi-equilibrium state (right column of Fig.~\ref{fig:dens_compact})  and oscillations of the downstream atmosphere  occur only at a small amplitude (Fig.~\ref{fig:radius_compact}). 

The downstream part of the remnant atmosphere is shielded from the ICM wind. Consequently, gas is not stripped directly from the downstream atmosphere. Instead, as the slow backflow in the near galactic wake encounters the downstream atmosphere, it slowly transports gas from the downstream atmosphere towards the sides of the atmosphere as shown in Fig.~\ref{fig:removal}. Only when gas is exposed to the ICM is it dragged along and finally stripped. As a result, much of the downstream atmosphere can survive up to or even beyond pericenter passage. Thus, the remnant atmosphere is distinctly asymmetric, extending farther downstream than upstream. 

The longer downstream part can have the visual appearance of a tail, especially for initially more extended galactic atmospheres (see Sect.~\ref{sec:extended_flow} and Fig.~\ref{fig:dens_extended}). We refer to this elongated unstripped downstream part of the atmosphere as the `remnant tail'. The gas fraction slices in Fig.~\ref{fig:overview} clearly show that the remnant tail consists almost exclusively of galactic gas, i.e., it is not mixed.  The  remnant tail is part of the `obstacle' the atmosphere poses to the ICM flow. Thus, the analogy to the flow past a solid body should not use a sphere but an elongated body.

\subsubsection{Evolution of the atmosphere after pericenter passage}

After pericenter passage, the ICM wind decreases and the upstream radius remains approximately constant. Gas removal along the sides continues, and finally the slow downstream stripping by the downstream backwards flow in the deadwater region  catches up and erodes the remnant tail {(last panels in Figs.~\ref{fig:dens_compact} and \ref{fig:dens_extended})}. Otherwise, flow patterns remain as described for the quasi-equilibrium phase above.

\subsubsection{Fate of the stripped gas}  \label{sec:flowaround}
 The wake of a stripped galaxy is of particular interest because galactic gas stripped from the atmosphere is transported into the wake and traces the wake closely. Here, stripped-off galactic gas could potentially mix with the ICM unless the mixing is prevented by ICM viscosity or  magnetic fields (see galactic gas fraction panels in Fig.~\ref{fig:overview}).   
In paper II we show that the global or average flow patterns discussed here also exist in viscous stripping,  but mixing in the wake is indeed suppressed (see Fig.~\ref{fig:overview}). 
  
Part of the stripped, or displaced,  galactic gas can also be trapped in the deadwater region in the wake for an extended time, in analogy to the dust behind a moving truck {(see Sect.~\ref{sec:hydro_steady})}. For example, we followed  the dark green tracer particles in Fig.~\ref{fig:removal} back to the start of the simulation. They originate from the downstream edge of the initial atmosphere, indicating that gas that is deposited in the deadwater region can linger there for at least several 100 Myr. 
  Thus mixing in the deadwater region will depend also on the amount of trapped galactic gas.   As described above and as evident from Fig.~\ref{fig:removal}, only gas from two regions is directly transported away from the galaxy: gas from the outer layers of the remnant atmosphere and wake, and gas from the far wake.  
 
In the context of spiral galaxy stripping, \citet{Schulz2001} and others described the fallback of stripped gas clumps due to the galaxy's gravity. This effect certainly plays a role for dense gas clumps which are less susceptible to ram pressure due to their small cross-section, they move almost  ballistically in the galaxy's potential. For stripping of diffuse gas described here the hydrodynamics effects in the wake dominate the motion of the stripped gas.

\subsubsection{Flow inside the atmosphere} \label{sec:innerflow}
%

\begin{figure}
\includegraphics[trim= 0 0 100 0,clip,angle=0,width=0.49\textwidth]{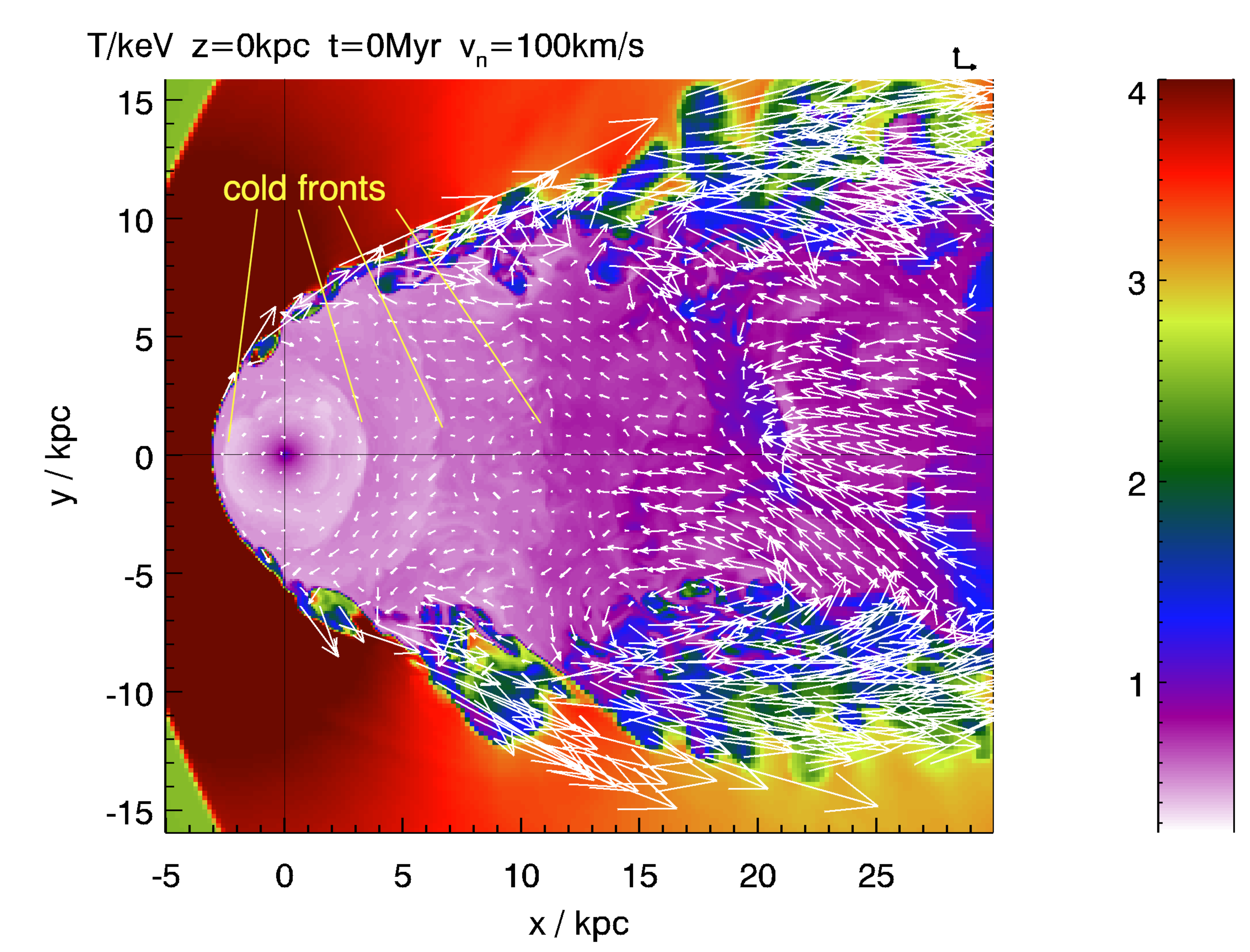}
\caption{Flow patterns inside the remnant atmosphere: Temperature slice through the atmosphere (initially extended atmosphere, inviscid) at pericenter passage. We overplot velocity vectors where the galactic gas fraction exceeds 0.3. Sloshing of the atmosphere leads to the marked sloshing cold fronts. Turbulent motions are superimposed to the sloshing.  {KHIs at the interface between the galactic gas and the ICM lead to vortex-like flows across the interface such as the one in the  bottom right    of this figure.}
\label{fig:innersloshing}}
\end{figure}


The galaxy's atmosphere is gaseous itself. Instead of the flow around a solid body, the scenario of a drop of heavier fluid falling through a lighter fluid could be a closer terrestrial analogue. At the sides of such a drop, either KHIs or viscosity transfer momentum into the drop, pulling  its outer layers  into the downstream direction. In turn, a vortex-like flow is established \textit{inside} the drop, such that the inner part of the drop moves towards the upstream interface. In the long run, the initially spherical drop turns into a cap-shape and a ring, which eventually breaks apart due to Rayleigh-Taylor instabilities (assuming sufficiently low surface tension).  

A similar pattern could be expected in the atmosphere of the stripped galaxy, but  never occurs in our simulations.  Instead, after the onset of the ICM wind, a vortex ring is formed in the \textit{downstream} galactic gas or in the downstream ICM (Figs.~\ref{fig:dens_compact} and \ref{fig:dens_extended}, left panels in second or first row, respectively). This vortex ring does not transport gas to the upstream edge of the atmosphere because its ram pressure is too weak to displace the dense central gas peak from the deep central potential well. For shallow central gravitational and gas density gradients this could be possible (e.g., in the simulations of \citealt{Heinz2003}), and may be observed in the cluster Abell 115 (\citealt{Forman2014}). 

Inside our model galaxies, the initial push by the onset of ICM flow leads to a series of long-lived sloshing motions of the atmosphere. {Such buoyancy-driven oscillations occur in any stratified atmosphere that is perturbed. For example, minor mergers of galaxy clusters can trigger sloshing in the atmosphere of the larger merger partner (\citealt{Markevitch2001,Ascasibar2006,Roediger2012a496}). In Fig.~\ref{fig:innersloshing} we show that the internal sloshing leads to a series of weak density and temperature discontinuities inside the atmosphere, which are sloshing cold fronts just like their counterparts seen in minor cluster mergers. The sloshing phenomenon is independent of the stripping process and simply superimposed on it.  }  The sloshing cold fronts in the upstream part of the atmosphere are stripped by the ICM wind, but they remain intact in the remnant tail. 
These internal sloshing motions are superimposed with  irregular motions of a few tens   $\Kms$.

\subsection{Initially compact vs. extended galactic atmosphere} \label{sec:extended_flow}
%

\begin{figure*}
\includegraphics[angle=0,width=0.99\textwidth]{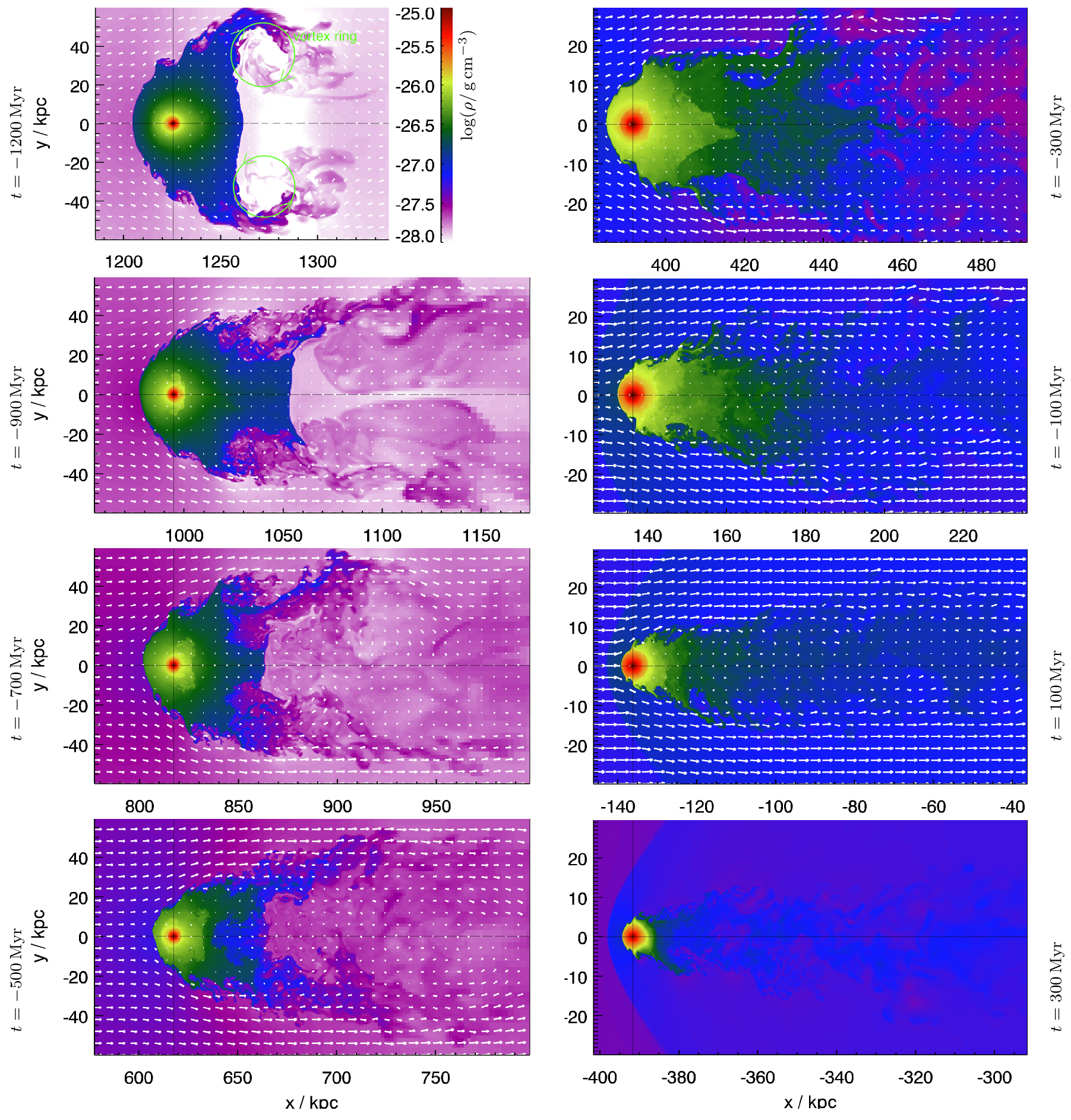}
\newline
\caption{Inviscid stripping of initially extended atmosphere (Sect.~\ref{sec:extended_flow}). 
Density slices through galaxy in orbital plane with velocity vectors overlaid in most panels. The left column shows the initial relaxation phase, the right column the quasi-equilibrium flow. The simulation was started 1.5 Gyr (1.4 Mpc) prior to pericenter passage. The initial oscillations last almost a Gyr. The galaxy retains much of its downstream atmosphere up to pericenter passage, giving rise to a pronounced remnant tail. Only $\sim 300 \Myr$ after pericenter passage is the remnant tail eroded.
\label{fig:dens_extended}}
\end{figure*}

\begin{figure}
\includegraphics[trim= 0 0 0 0,clip,angle=0,width=0.45\textwidth]{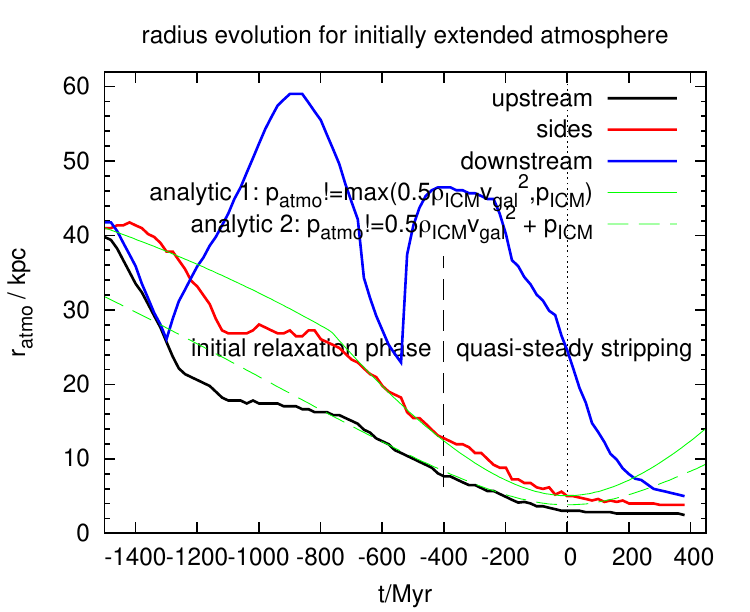}
\includegraphics[trim= 0 0 0 15,clip,angle=0,width=0.45\textwidth]{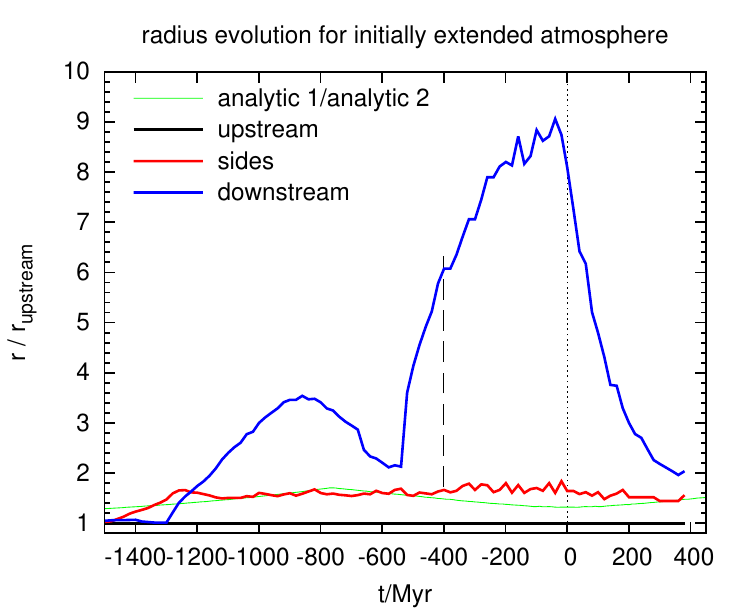}
\caption{Evolution of the size of the remnant atmosphere towards the upstream and downstream direction and towards the sides for the initially extended atmosphere (see Fig.~\ref{fig:dens_extended} for snapshots). Otherwise same as Fig.~\ref{fig:radius_compact}. Note the stronger downstream oscillations and the longer extent of the downstream atmosphere.
\label{fig:radius_ext}}
\end{figure}


Figure~\ref{fig:dens_extended} displays snapshots of the stripping of an initially extended atmosphere (galactic gas density $\propto r^{-1.2}$ instead of $\propto r^{-1.5}$). As in Fig.~\ref{fig:dens_compact}, the left and right columns show the initial relaxation phase and the quasi-equilibrium phase, respectively. Fig.~\ref{fig:radius_ext} summarizes the size evolution of this atmosphere. The bottom panel of Figure~\ref{fig:vanDykeonly} shows a labelled density slice from this simulation for direct comparison to the compact atmosphere and the flow around a solid sphere.  

Qualitatively the stripping proceeds in the same manner as described above.  A more extended initial atmosphere leads to a longer relaxation phase after flow onset; in the case shown here it lasts almost 1 Gyr. As we simulated this galaxy on the orbit with a faster pericenter velocity, this simulation started 1.5 Gyr prior to pericenter passage to ensure the flow onset is subsonic. Also here, as in our compact atmosphere case,  stripping reaches a quasi-equilibrium about 0.5 Gyr prior to pericenter passage. 

Second, an initially more extended atmosphere retains a longer downstream atmosphere -- or remnant tail -- for a longer time because, as explained above, the downstream atmosphere is shielded from the ICM wind and downstream stripping is very slow. In the case presented here a clear remnant tail survives until 300 Myr (390 kpc) after pericenter passage, and it can be  6 to 9 times longer than the upstream radius.  Despite this  length, the remnant tail contains unstripped, unmixed galactic gas. 

For either atmosphere, the deadwater region starts  downstream of the remnant tail and is typically 1.5 times as long as the {total length} of the complete remnant atmosphere, i.e.,  including the remnant tail.  For the extended atmosphere,  turbulence in the deadwater region and slow turbulence inside the remnant tail somewhat blur the downstream contact discontinuity between these two flow regions. The combination of the longer remnant tail, the longer deadwater region, and the initially more extended galactic atmosphere lead to more galactic gas being trapped in the deadwater region.

\section{Discussion} \label{sec:discussion}
Our simulations reproduce the basic features of elliptical galaxy stripping found by previous authors, i.e., the bow shock, upstream contact discontinuity and downstream {`tail', in the broadest sense} (e.g., \citealt{Acreman2003,Stevens1999,Toniazzo2001}). Going beyond previous work, we linked the flow patterns arising in galaxy stripping to the flows around solid bodies. In this course, we distinguished the remnant tail of the atmosphere from the galaxy's wake, which had been combined into the term `tail' before. These flow patterns can be seen in snapshots presented in earlier work (e.g., \citealt{Acreman2003,Stevens1999}), but their origin and implications have, to our knowledge, not been discussed.  Distinguishing between the remnant tail and the wake is important because only in the latter can mixing between stripped galactic gas and ICM  take place. The remnant tail always contains pure galactic gas only, and thus cannot be used as a tracer of ongoing or suppressed mixing.  \citet{Shin2013} argue that a sufficient level of turbulence inside the galactic atmosphere can lead to diluting the galactic atmosphere with ICM over a timescale of Gyrs. However, their study does not include a realistic stripping history. Given the long dilution timescale, the effect might be small in a realistic case.

Furthermore, we discussed the origin and duration of the flow initialization phase where stripping has not yet reached a quasi-steady state, and the tail and wake structure is still dominated by initial conditions. 
We demonstrated the long-lasting impact of  the {initial  extent of the galactic atmosphere}, which lasts well into the quasi-equilibrium stripping phase. Consequently,  direct comparisons between generic simulations and  observations of particular galaxies can be difficult and misleading. In particular, simulations that expose model galaxies or subclusters to a constant ambient wind cannot reach a quasi-equilibrium stripping representative of an ongoing infall into the cluster. First, such a model galaxy experiences the initial relaxation phase where the downstream atmosphere oscillates, and then directly enters a stage comparable to the after-pericenter stripping where the ambient wind does not increase further.  For example, the simulations of \citet{Suzuki2013} and \citet{Zavala2012} cover the initial relaxation phase only.  {Due to the missing phase of gradual increase of the ICM wind  the structure of the remaining atmosphere and the wake are different. For example, the remnant tail will not be as pronounced as in simulations that cover a full infall. Furthermore, the sudden onset of a strong ICM wind leads to strong sloshing motions of the remnant atmosphere that enhance the gas loss.} 

The long-lasting signatures of the initial atmosphere's extent combined with the long-lasting flow relaxation phase could be especially interesting for groups or galaxies in an early phase of their infall into their host cluster. Such examples include M49 in the southern outskirts of the Virgo cluster, and groups in the peripheries of the clusters Abell 2142 (\citealt{Eckert2014a}) and Abell 85 (\citealt{Kempner2002}). If these groups have passed the accretion shock just recently, their appearance is likely still  dominated by irregular flow patterns typical for the initial relaxation phase. For the same reasons, large galaxies falling into small clusters or groups may never reach the quasi-equilibrium phase during their first infall because in the more compact cluster atmosphere the ram pressure changes more rapidly than the galaxy can adapt to.

\section{Summary} \label{sec:summary}
We determined the similarities and differences between flows around a solid blunt body on the one hand and flows in gas-stripping of elliptical cluster galaxies on the other hand. To this end, we simulated gas-stripping of the hot atmosphere of the Virgo cluster elliptical M89 (NGC 4552), taking into account a realistic gravitational potential, initial gas contents, and orbit through the Virgo cluster. 

First, we distinguish two aspects of the stripping process:
\begin{itemize}
\item Evolution of the atmosphere: The atmosphere of an infalling spherical galaxy is {displaced somewhat} towards the downstream direction and is stripped preferentially along its sides. The downstream part of the atmosphere is shielded from the ICM wind, and a portion of the downstream atmosphere significantly larger than at the upstream part can survive stripping up to or beyond pericenter passage. This unstripped downstream gas forms the remnant tail. Only at or after pericenter passage is this remnant tail eventually stripped.
\item Flow around the atmosphere: The flow of the  ICM \textit{around} the evolving atmosphere  resembles the flow around a solid body.  However, the equivalent of the solid body is not a sphere or ellipse, but the complete remnant atmosphere including the remnant tail. 
The \textit{wake} starts only downstream of  the remnant tail. The first part of the wake is a deadwater region. Gas stripped from the galaxy traces the deadwater region and far wake. 
\end{itemize}
The  length and density of the remnant tail and the deadwater region increase for an initially more extended galactic atmosphere.  {Mixing between galactic gas and ambient ICM begins only in the deadwater region and increases along the wake, but does not occur in the remnant tail. }Thus, observations aiming at determining ICM plasma properties via progressing or suppressed mixing between galactic gas and the ICM must probe the \textit{wake} sufficiently downstream of the galaxy center to avoid confusion with the remnant tail.  

Second, we demonstrated that a sudden onset of stripping is followed by an extended initial relaxation phase, whose presence is also known from flows past solid bodies. Galaxies or groups being stripped in cluster outskirts likely are still in the initial relaxation phase. Only galaxies nearing their pericenter passage have reached a quasi-steady stripping phase.

Paper II of this series investigates the impact of ICM viscosity on the flow patterns and derives observable signatures of inviscid and viscous stripping. Paper III gives a  detailed comparison of archival XMM and new Chandra data to the simulation results.

\acknowledgments
The FLASH code was in part developed by the DOE NNSA- ASC OASCR Flash center at the University of Chicago. E.R.~acknowledges the support of the Priority Programme Physics of the ISM of the DFG (German Research Foundation),  the supercomputing grants NIC 6006 and 6970 at the John-Neumann Institut at the Forschungszentrum J\"ulich,  a visiting scientist fellowship of the Smithsonian Astrophysical Observatory, and the hospitality of the Center for Astrophysics in Cambridge. We are grateful for helpful discussions with Marcus Br\"uggen and Dominique Eckert. We also thank the referee for helpful comments regarding the presentation of our results.
This research has made use of the GOLDMine Database.



{\it Facilities:} \facility{CXO (ACIS)}.

%
\bibliographystyle{apj}
\bibliography{library}


\appendix

\section{Tailoring the simulations to the Virgo elliptical M89} \label{sec:tailor_m89}

\subsection{Galaxy potential and initial gas contents} \label{sec:galaxymodel}
In the current work we neglect the ellipticity of M89 and model this galaxy as  a spherical gravitational potential filled with an initially hydrostatic atmosphere. Stars and dark matter (DM) dominate the galaxy's mass, hence we neglect self-gravity in the gas and use a static potential. For simplicity we describe the galactic potential as a superposition of three Hernquist models (\citealt{Hernquist1990}).  Hernquist potentials have the advantage of a finite total mass. The gravitational potential $\Phi(r)$ and cumulative mass $M(r)$ of component $i$ depend on the corresponding total mass $M_i$, scale radius $a_i$, and circular radius $r$ as
\begin{eqnarray}
\Phi(r)  &=& - G \frac{M_i}{r+a_i} , \\
 M(r) &=& M_i \frac{r^2}{(r+a_i)^2}.
\end{eqnarray}

\begin{figure*}
\includegraphics[width=0.49\textwidth]{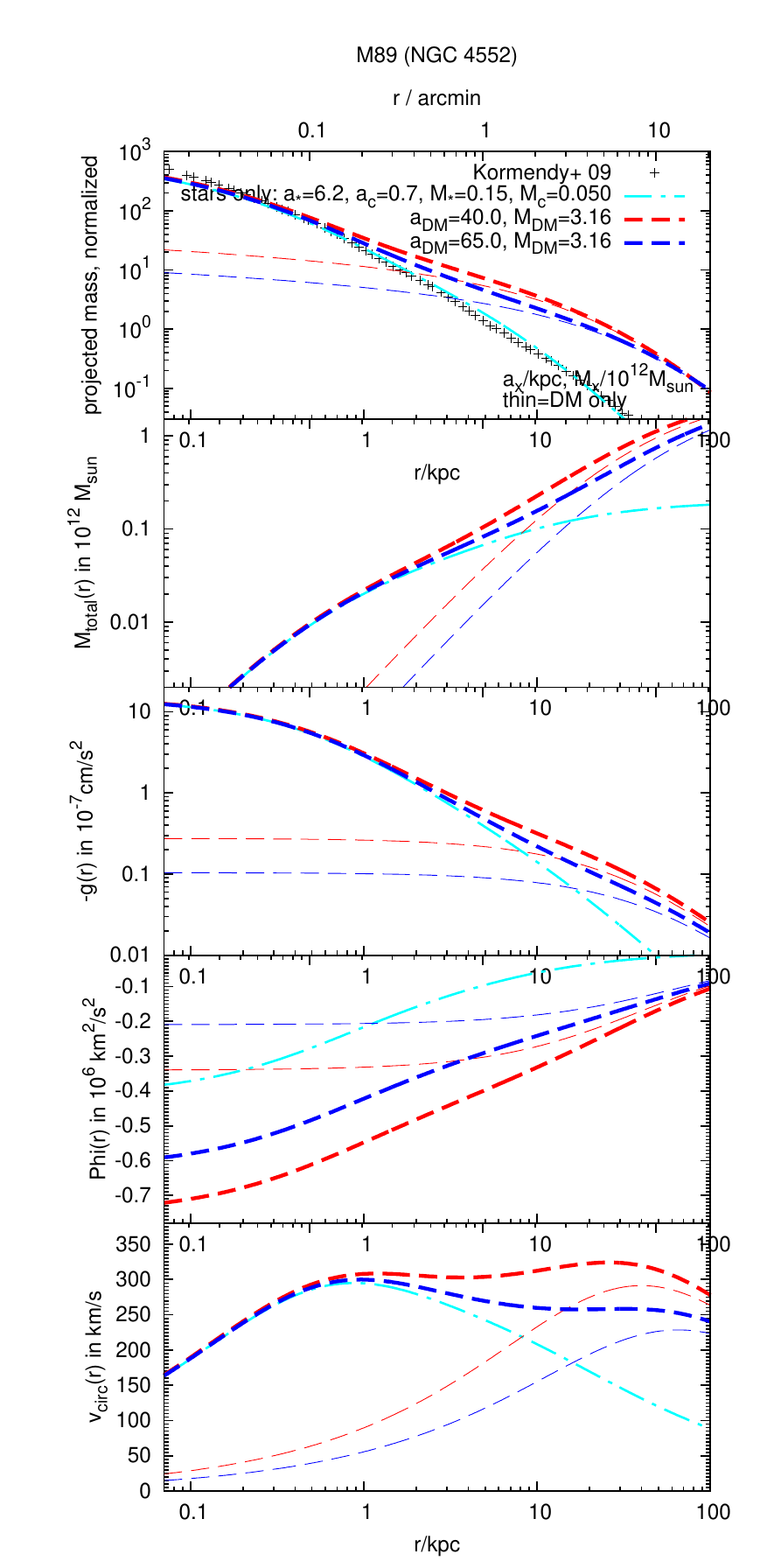}
\includegraphics[width=0.49\textwidth]{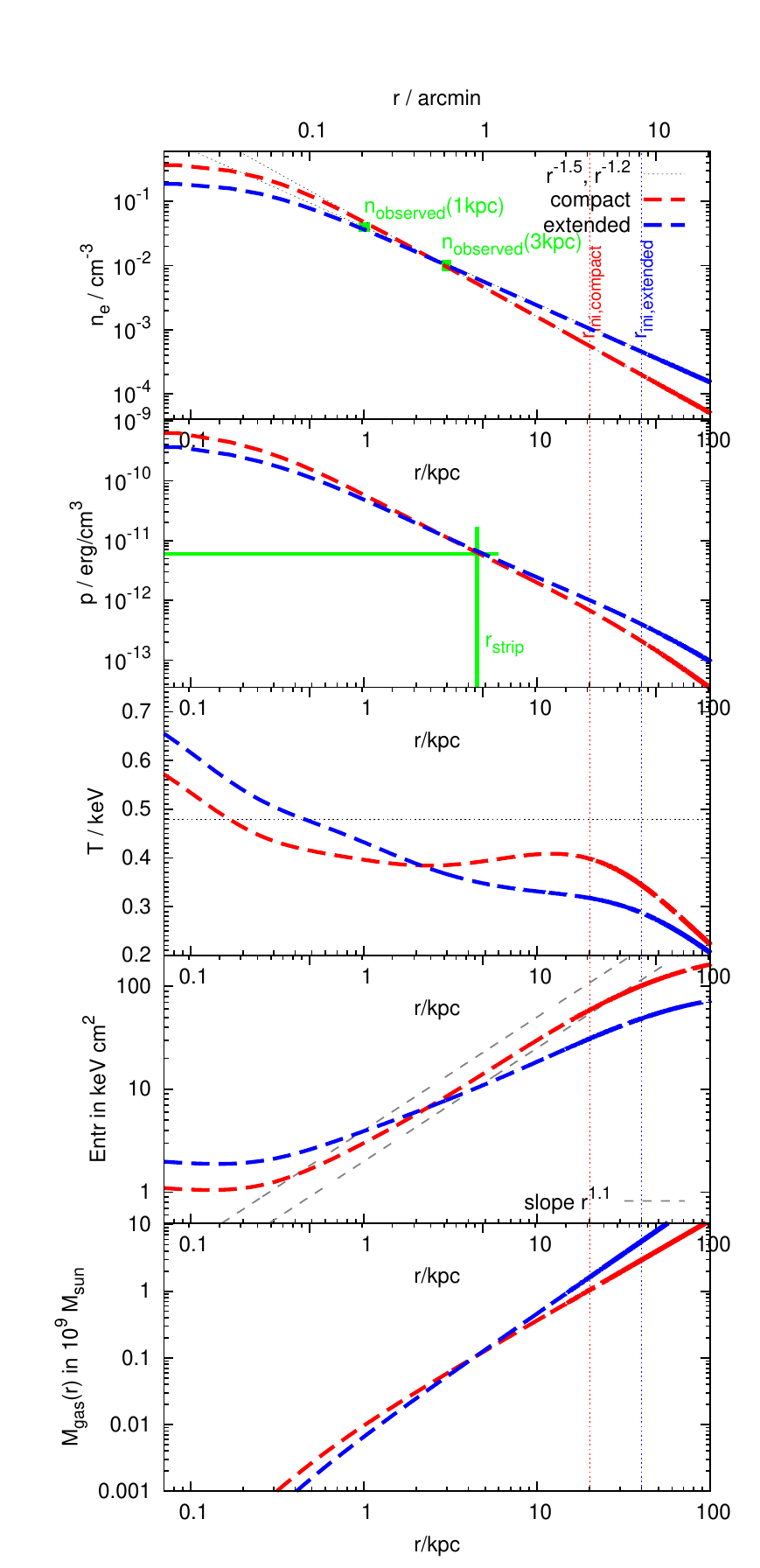}
\caption{Model galaxy potentials (left) and galactic gas profiles (right). In the left column we plot radial profiles of projected mass and comparison to brightness profile, cumulative mass, gravitational acceleration, gravitational potential, and circular velocities. The thin lines are for the DM component only. In the right column we plot the corresponding radial profiles of electron density, thermal pressure, ICM temperature, entropy, and cumulative mass profiles. In green we mark observational constraints on the galactic atmosphere. See also App.~\ref{sec:galaxymodel}. Note that the gas profiles are relevant only out to the initial radius of the atmosphere as set by pressure balance with the initially ambient ICM (marked by thin vertical lines of matching color).}
\label{fig:galaxyprofs_M89}
\end{figure*}

The inner potential is known from its stellar light, the inner hot gaseous atmosphere from the Chandra observation of the remnant atmosphere. The outer potential and initial atmosphere are not directly known because M89 is already heavily gas-stripped; thus we use results for comparable non-stripped ellipticals and encompass the reasonable parameter space for M89 in two models. Figure~\ref{fig:galaxyprofs_M89} summarizes mass, potential, and gas profiles of the initial models. Table~\ref{tab:M89model} lists the parameters of the galaxy models.

%

\begin{table}
\caption{Galaxy model parameters for M89.}
stellar distribution common parameters to all models:
\newline
\begin{tabular}{ll}
\tableline\tableline
$M_*/ 10^{11}M\subsun$ & 1.5\\
$a_* / \Kpc$ & 6.2 \\
$M_c/ 10^{11}M\subsun$ & 0.5\\
$a_c / \Kpc$ & 0.7 \\
\tableline
\end{tabular}
\newline
\newline
\newline
\begin{tabular}{lll}
\tableline\tableline
\multicolumn{3}{l}{DM halos and galactic atmospheres:}\\
                                           & compact & extended \\
                                           \tableline
$M\DM / 10^{12}M\subsun$ & 3.16 & 3.16 \\ 
$a\DM / \Kpc$ & 40. & 65. \\ 
$n_0 / \ccm$ & 0.4  & 0.2\\
$r_c / \Kpc$  & 0.25 & 0.25 \\
$\beta$          & 0.5 & 0.4\\
truncation radius/kpc & 19 & 43 \\
\tableline
\end{tabular}
\label{tab:M89model}
\end{table}%

\subsubsection{Stellar component}
We approximate M89's stellar component with a double Hernquist model, which fits its stellar light profile (\citealt{Kormendy2009}) between 0.1 and 30 kpc (top left panel of Fig.~\ref{fig:galaxyprofs_M89}). Scale radii and masses for the main and core stellar component are given in Table~\ref{tab:M89model}. There are several measures of the total stellar mass of M89. \citet{Zibetti2011} convert the total H band luminosity of M89 to a total stellar mass of $10^{11}M\subsun$. \citet{Cappellari2006} and \citet{Emsellem2004} give stellar velocity dispersions of  $250\Kms$, which corresponds to a virial mass of $1.5\times 10^{11}M\subsun$ within 10 kpc. The gaseous atmosphere of M89 is  disturbed by gas stripping and a central AGN outburst. The measured gas temperature in the remnant gas core excluding the central AGN bubbles is $0.48\pm 0.03$ keV (paper III). Assuming that this value is representative for the undisturbed  atmosphere, this temperature corresponds to a virial mass of $3\times 10^{11}M\subsun$ within 10 kpc radius. 

A choice of $M_*=1.5\times 10^{11}M\subsun$ and $M_c=0.5\times 10^{11}M\subsun$ yields a galactic gas temperature of 0.4 keV around $ 3\Kpc$, which is still 20\% below the observed temperature. However, the optical data does not support an even higher stellar mass. A lower  stellar mass would reduce the gas temperature even further. M89 is currently experiencing an AGN outburst and may have been heated somewhat by earlier outbursts. Thus, the temperature of the model atmospheres agrees reasonably with the observations.

\subsubsection{Gaseous atmosphere}
Only the central part of the hot atmosphere of M89 remains, hence we extrapolate this atmosphere based on observations of other elliptical galaxies. We describe the galactic gas density profile by a $\beta$ model, using $\beta=0.5$ for a compact atmosphere (comparable to NGC 1404, \citealt{Scharf2005}) and $\beta=0.4$ for a more extended atmosphere more typical for galaxy-size and group-size DM halos. We normalize the density profile to agree with the power-law fit by  \citet{Machacek2006a} between 1 and 3 kpc (see top right panel in Fig.~\ref{fig:galaxyprofs_M89}). 

We assume the initial atmosphere to be in hydrostatic equilibrium and compute the pressure and temperature profiles according to total gravitational potential. For a chosen  galactic gas density, the DM potential sets the galactic gas temperature and pressure profile as described below. We assume a zero pressure at infinity and integrate the galactic atmosphere pressure profile inwards. The model galaxy is then placed at its initial position in the cluster outskirts, and its atmosphere truncated where its pressures falls below the ambient ICM pressure. The truncation radius for both models is listed in Tab.~\ref{tab:M89model} and marked in Fig.~\ref{fig:galaxyprofs_M89}.

\subsubsection{Dark matter (DM) halo}
 The total mass of elliptical galaxies is dominated by DM albeit with considerable scatter (e.g., review by \citealt{Napolitano2012}). For example, \citet{Das2010} measure DM fractions of  about 80 to 90\% within 4 effective radii. Thus, we expect a DM mass  of roughly $2 \times 10^{12}M\subsun$ in M89. \citet{Humphrey2011,Humphrey2012} report mass models for the elliptical galaxies NGC 720 and NGC 1521, which are comparable to M89 in stellar mass and gas temperature but still have  extended gaseous atmospheres. They find the following characteristics:  NGC 720 -- stellar mass of $10^{11}M\subsun$, total mass of $10^{12}M\subsun$ inside 70 kpc, cumulative stellar and DM mass are equal at a radius of 6 kpc, gas temperature is 0.5 keV, declining outside 10 kpc; NGC 1521 -- stellar mass of  $4\times 10^{11}M\subsun$,  total mass of  $2\times 10^{12}M\subsun$ inside 100 kpc and  $5\times 10^{12}M\subsun$ inside 200 kpc, stellar mass exceeds DM mass out to 18 kpc, gas temperature about 0.5 keV, rising up to 0.8 keV outside 20 kpc.

Based on these constraints we experimented with DM halos between 1 and 10$\times 10^{12}M\subsun$ and found the main characteristics of the stripping independent of the exact choice. In this work we use a total DM mass of $3.16\times 10^{12}M\subsun$. We chose the scale radii for the DM halos for our galaxy models with the compact and extended atmosphere case (see Table~\ref{tab:M89model}) such that we obtain an approximately flat circular velocity and  gas temperature profile out to $\sim 40\Kpc$, and an entropy slope  comparable to the results for NGC 720 and NGC 1521. The cumulative gravitating mass and the gravitational acceleration in the central 3 kpc are almost identical between our models, and the pressure profiles agree within 25\% between 1 and 8 kpc from the galaxy center.

\subsection{Galaxy orbit/ICM wind}  \label{sec:ICMwind}

\subsubsection{Observational constraints} \label{sec:ICMwind_obs}
M89 is currently located 350 kpc (72 arcmin) east of M87 in projection. The position of M89 along our LOS inside the Virgo cluster is uncertain. \citet{Tonry2001} place it at the same distance as M87, but surface brightness fluctuation measurements by \citet{Mei2007} suggest M89 is about 1 Mpc closer to us than the Virgo center. Also from surface brightness fluctuations, \citet{Blakeslee2009} list a distance of  $16\pm 0.5\Mpc$, which is  0.7 Mpc closer than their listed M87 distance ($16.7\pm 0.6\Mpc$), but consistent with the distance of M87 within the error bars. 

M89 is moving towards us at $725\Kms$ or $967\Kms$ with respect to the Virgo mean or M87, respectively.
This is close to the Virgo ICM sound speed. Its gas tail to the south and its contact discontinuity to the north suggest a signifiant velocity in the plane of the sky towards the north. \citet{Machacek2006a} derive an absolute velocity of M89 through the Virgo ICM of $1700\Kms$ (Mach 2.2), based on its stagnation point pressure and assuming M89 and the Virgo center have the same distance to us. This velocity is rather high. Assuming M89 was at a larger real-space distance to the Virgo center (M87) would increase the total velocity even more, because the ambient reference pressure would be even lower, and the measured high stagnation pressure could be achieved only by a higher velocity. A velocity higher than Mach 2 at a cluster-centric distance of $\gg 300\Kpc$ is unlikely, hence we assume that M89's projected cluster-centric distance is close to the real-space distance.  Given its projected northward motion and location straight east of the cluster center, M89 is most likely just passing the pericenter of its orbit through Virgo. This geometry is supported by our comparison between observations and simulations (paper III). If M89 was indeed 1 Mpc closer to us than the Virgo center and still moving towards us, it would have passed its pericenter many 100 Myr ago, and should have lost its remnant tail. The presence of the long remnant tail in M89 thus argues against this scenario, and instead places M89 close to pericenter passage.

 With a real space velocity of $1700\Kms$ and a radial velocity of $725\Kms$ towards us, M89's orbit would be inclined by 25 degree out of the plane of the sky. Machacek et al.'s estimate of the total velocity does not include effects of ICM falling into the M89 potential from the upstream region and may thus somewhat overestimate the orbital velocity.   Even if the true orbital velocity would be as low as $1000\Kms$, the orbit would be inclined out of the plane of the sky by only 45 degree.

\subsubsection{ICM wind in simulation box}

For simplicity we run our simulations in the rest frame of the galaxy, i.e., the galaxy is exposed to an ICM flow or wind. This wind has a constant direction but varies in density and velocity according to the assumed motion of the galaxy through the Virgo cluster. Varying the ICM wind requires  shifting the cluster potential through the simulation grid to account for the non-inertial rest frame.  We neglect gradients in the ICM and cluster potential perpendicular to the orbit, i.e., we approximate the galaxy's motion through the cluster by a `straight wind tunnel'. This is a good approximation as long as the galaxy does not pass close to the cluster center, which is the case for M89.  

\begin{table}
\caption{ICM parameters for the Virgo cluster model, updated version from \citet{Roediger2011}.}
\begin{tabular}{lcc}
\tableline\tableline
density: & \multicolumn{2}{c}{double $\beta$ profile} \\
\tableline
core radii $r_{1,2}/\Kpc$: & 1.7 & 21.4\\
core densities $\rho_{01,02}/(\gccm)$:  & $3.38 \cdot 10^{-25}$ & $1.175\cdot 10^{-26}$\\
$\beta_{1,2}$: & 0.42 & 0.47\\ 
\tableline\tableline
temperature: & \multicolumn{2}{c}{ see Eqn.~1 in \citet{Roediger2011}}\\
\tableline
slopes $m_{1,2}/(\K \PC^{-1})$: & 0. &  0.\\
$T_{01,02}/(\KeV)$: & 1.53 & 2.4\\
break radius $r\Break/\Kpc$: & \multicolumn{2}{c}{ 22.6 }\\
break range $a\Break/\Kpc$: & \multicolumn{2}{c}{ 6.7 }\\
\tableline
\end{tabular}
\label{tab:Virgo_parameters}
\end{table}%

\begin{figure}
\includegraphics[width=0.45\textwidth]{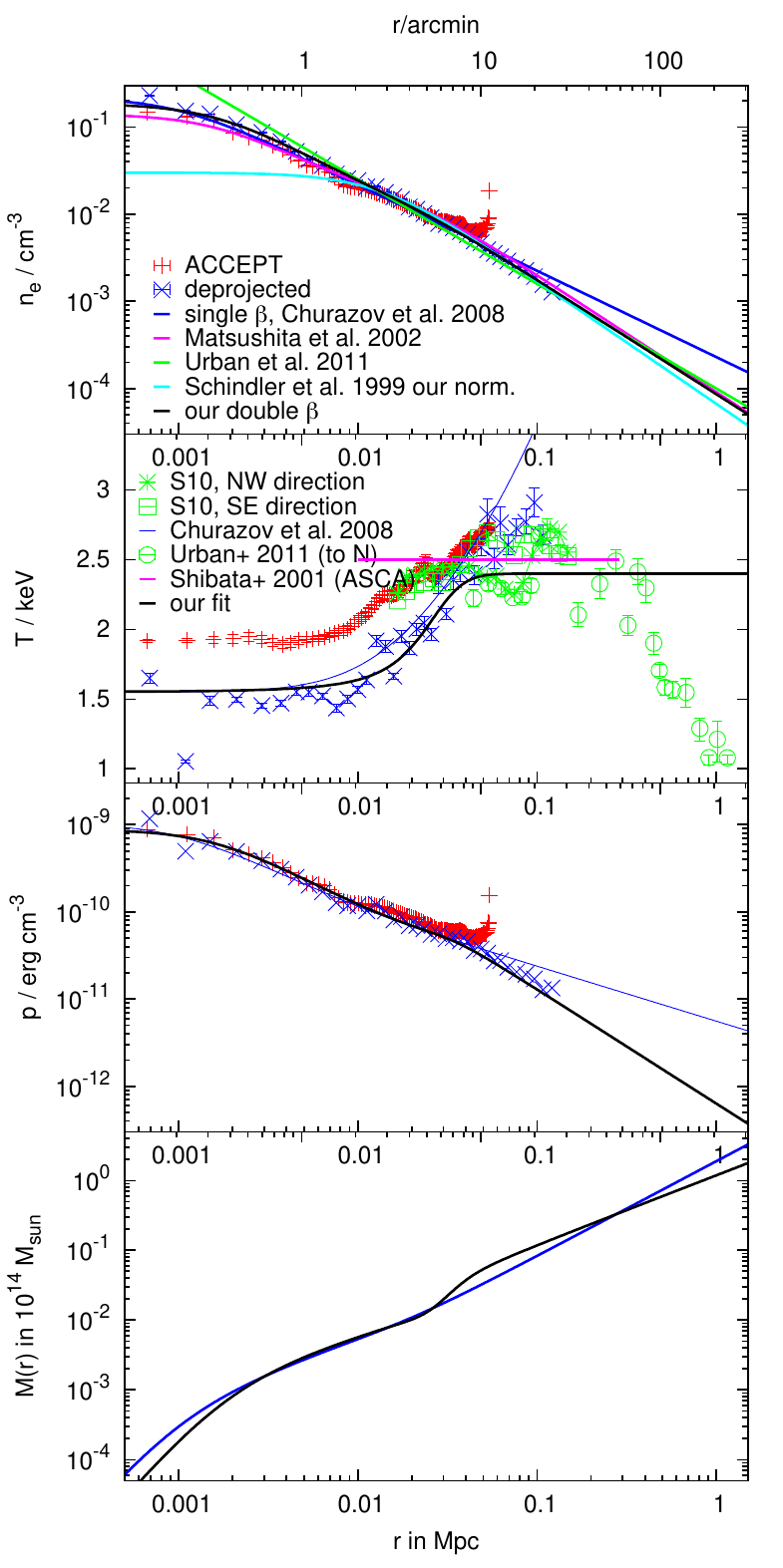}
\caption{Virgo cluster radial profiles, from top to bottom: (i) Electron density, $n_e$, from the ACCEPT sample (\citealt{Cavagnolo2009}) and the deprojected profile,  along with a single- or double-$\beta$ model fits from \citet{Churazov2008}, \citet{Matsushita2002}, \citet{Urban2011}, \citet{Schindler1999}, and our fit. (ii) Temperature, from \citet{Simionescu2010} in two sectors, \citet{Churazov2008}, \citet{Urban2011}, \citet{Shibata2001}, and our fit. (iii) Resulting pressure profile for selected data sets/fits, and (iv) corresponding cumulative mass profile for the Virgo cluster.}
\label{fig:Clusterprofs}
\end{figure}

We calculate the variable ICM wind as follows: We set up a spherical model for the Virgo cluster. Radial ICM density and temperature profiles are described by a double beta model and Eqn.~1 in \citet{Roediger2011}, respectively. Table~\ref{tab:Virgo_parameters} lists our updated parameters, now taking into account the measurements out to the Virgo virial radius by \citet{Urban2011}. We neglect the temperature decrease beyond 400 kpc in order to keep a constant dynamic viscosity during infall in our simulations. {Thus, we over-estimate the viscosity on the first part of the orbit, such that during the early infall smaller KHIs than in our simulation could grow. However, the highest viscosity we used was only 10\% of the Spitzer viscosity. Using instead a full Spitzer viscosity along the early orbit and a lower ICM temperature would roughly compensate each other, leading to similar results as in our viscous simulation.}
Figure~\ref{fig:Clusterprofs} summarizes available observational data and our fits. We assume the Virgo ICM is in hydrostatic equilibrium and calculate the corresponding radial gravitational acceleration and mass. 

\begin{figure}
\includegraphics[width=0.49\textwidth]{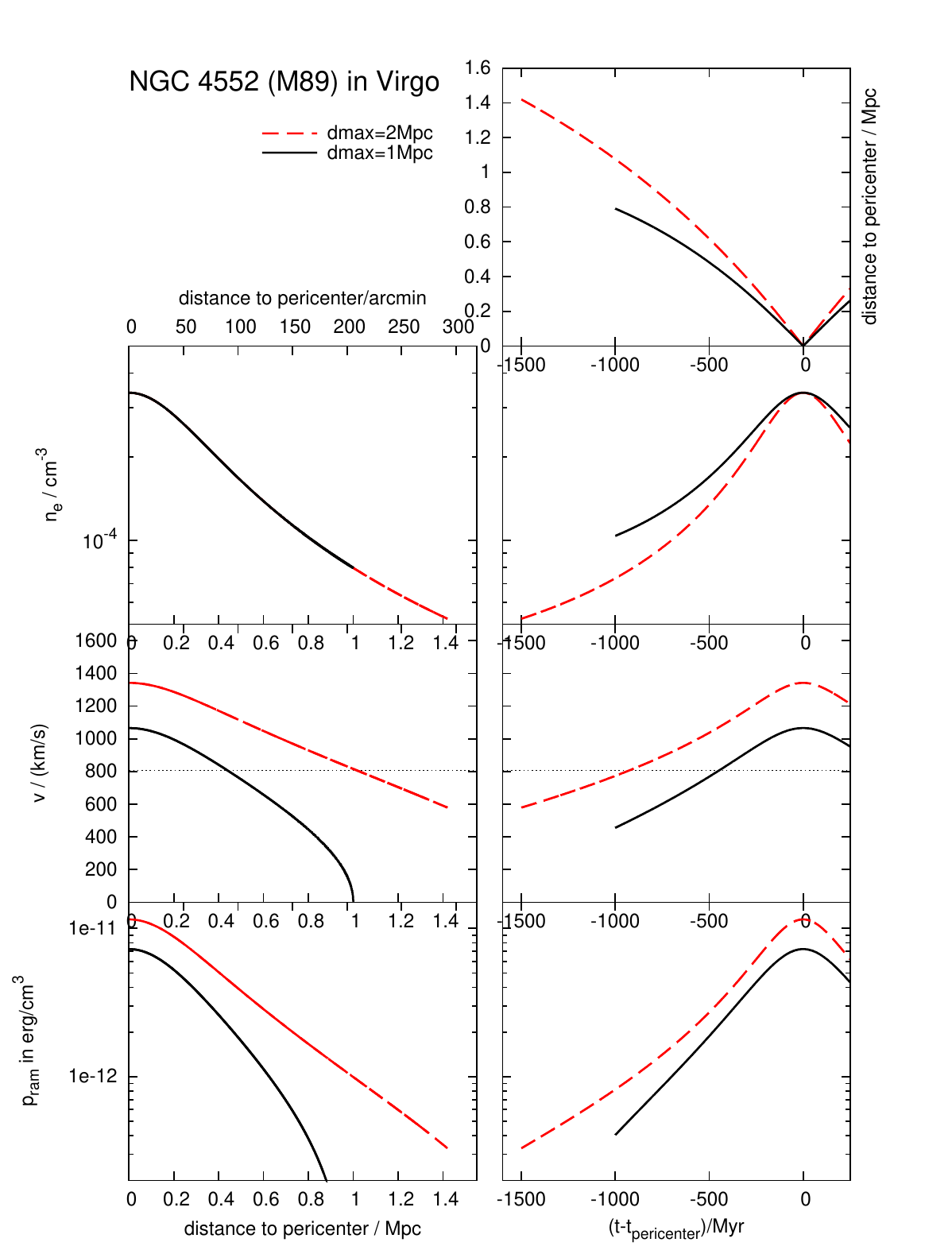}
\caption{ICM properties along orbits, as function of position and as function of time.}
\label{fig:orbits}
\end{figure}


We approximate the orbit of M89 by a test particle orbit in the Virgo potential. Such test particle orbits  curve only slightly at pericenter passage, hence we use the straight wind tunnel approximation as explained above. The pericenter distance is 350 kpc as observed for M89. We consider two orbits with apocenter distances of 1 and 2 Mpc and refer to them as the slow and fast orbit, respectively. The variation of ICM density, velocity, and ram pressure along these orbits is shown in Fig.~\ref{fig:orbits}. On these orbits, M89 reaches pericenter velocities of $1050\Kms$ and $1350\Kms$, respectively. Both are below the estimated total velocity despite the large apocenter distances. This again suggests that the velocity from the stagnation point method is somewhat overestimated due to the upstream attraction of ICM into the M89 potential. 

We want to ensure that the simulated galaxy reaches a quasi-equilibrium state and flow patterns are not dominated by initial conditions. Expecting an extended relaxation phase as explained in Sect.~\ref{sec:flowpatterns}, we  initialize the galaxy and its motion a generously long time and distance away from the pericenter ($\ge 1\Gyr$, $\ge 0.8\Mpc$). The gentlest initialization would be to start the infall from zero velocity.  With much simulation time spent in the far outskirts of the cluster it would, however,  be prohibitively expensive. Therefore we start our simulations with a subsonic ICM velocity of around Mach 0.6. The initial ICM flow is set as a potential flow past the initially spherical  galactic atmosphere (first panel in Fig.~\ref{fig:dens_compact}). The abrupt but subsonic onset of the flow is probably even more realistic than a fully smooth start, because in cluster outskirts subsonic large-scale turbulence or bulk flows are expected to inflict abrupt changes in the ICM wind working on the infalling galaxy.

\end{document}